\documentclass[a4paper,twoside,11pt]{article}
\usepackage{mathtools} 
\usepackage{bbm}
\usepackage{amsmath}%
\usepackage{amsfonts}%
\usepackage{amssymb}%
\usepackage{graphicx}
\usepackage[ruled]{algorithm2e}
\usepackage{a4wide}
\usepackage{subfigure}
\usepackage{hyperref}
\usepackage{geometry}
\usepackage{tikz}
\usetikzlibrary{arrows,shapes}
\usetikzlibrary{trees}
\usetikzlibrary{matrix,arrows} 				
\usetikzlibrary{positioning}				
\usetikzlibrary{calc,through}				
\usetikzlibrary{decorations.pathreplacing}  
\usetikzlibrary{decorations.pathmorphing}	
\usetikzlibrary{decorations.markings}
\tikzset{
    vector/.style={decorate, decoration={snake}, draw},
	provector/.style={decorate, decoration={snake,amplitude=2.5pt}, draw},
	antivector/.style={decorate, decoration={snake,amplitude=-2.5pt}, draw},
    fermion/.style={draw=black, postaction={decorate},
        decoration={markings,mark=at position .55 with {\arrow[draw=black]{>}}}},
    fermionbar/.style={draw=black, postaction={decorate},
        decoration={markings,mark=at position .55 with {\arrow[draw=black]{<}}}},
    fermionnoarrow/.style={draw=black},
    gluon/.style={decorate, draw=black,
        decoration={coil,amplitude=4pt, segment length=5pt}},				
    scalar/.style={dashed,draw=black, postaction={decorate},
        decoration={markings,mark=at position .55 with {\arrow[draw=black]{>}}}},
    scalarbar/.style={dashed,draw=black, postaction={decorate},
        decoration={markings,mark=at position .55 with {\arrow[draw=black]{<}}}},
    scalarnoarrow/.style={dashed,draw=black},
    electron/.style={draw=black, postaction={decorate},
        decoration={markings,mark=at position .55 with {\arrow[draw=black]{>}}}},
	bigvector/.style={decorate, decoration={snake,amplitude=4pt}, draw},
}

\begin{document}
\begin{titlepage}
\begin{flushright}
IFT-UAM/CSIC-14-004\\
FT-UAM-14-4\\[10pt]
\end{flushright}
\vspace{15truemm}
\begin{center}
  {\Large\bf Variance reduction with practical all-to-all lattice propagators\\[0.5ex]}
\end{center}
\vspace{10truemm}
\begin{center}
{\large E. Endress,$^{1}$ C. Pena,$^{1,2}$ and K. Sivalingam$^{3*}$
}
\vskip 0.5cm
$^{1}$Instituto de F\'{\i}sica Te\'orica UAM/CSIC,
 Universidad Aut\'onoma de Madrid,\\
  Cantoblanco E-28049 Madrid, Spain
\vskip 1.5ex
$^{2}$Dpto. de F\'{\i}sica Te\'orica, 
Universidad Aut\'onoma de Madrid,\\
Cantoblanco E-28049 Madrid, Spain
\vskip 1.5ex
$^{3}$Department of Meteorology,
University of Reading,\\
Earley Gate, PO~Box~43, Reading, RG6 6BB, UK
\vspace{20truemm}
\end{center}
{\small
{\bf Abstract: }
We discuss all-to-all quark propagator techniques in two (related) contexts within Lattice QCD:
the computation of closed quark propagators, and applications to the so-called ``eye diagrams''
appearing in the computation of non-leptonic kaon decay amplitudes.
Combinations of low-mode averaging and diluted stochastic volume sources that yield
optimal signal-to-noise ratios for the latter problem are developed.
We also apply a recently proposed probing algorithm to compute directly the diagonal
of the inverse Dirac operator, and compare its performance with that of stochastic methods.
At fixed computational cost the two procedures yield comparable signal-to-noise ratios,
but probing has practical advantages which make it a promising tool for a wide range
of applications in Lattice QCD.
}
\vfill
\noindent
$^*$e-mail: {eric.endress@uam.es, carlos.pena@uam.es, k.sivalingam@reading.ac.uk}
\end{titlepage}

\section{Introduction}
The computation of the diagonal of the inverse of a large, sparse matrix is
a demanding computational task.
One context where this problem has to be faced is the study of the low-energy
dynamics of strongly interacting elementary particles within Quantum Chromodynamics,
using the Lattice QCD (LQCD) approach.
In LQCD hadronic properties such as particle masses and matrix elements
can be computed from first principles in terms of Euclidean correlation functions.
The latter are expectation values, computed within the path-integral formulation
of Quantum Field Theory, of products of composite operators, made up from quark and gluon fields.
After integration over the quark fields, the correlation functions are expressed as traces over products of quark propagators and constant spin, color, and flavor matrices.
Quark propagators are in turn computed by inverting the lattice Dirac operator $D$, which within the
lattice approach is a large, sparse, complex matrix of typical dimension in the $10^6$-$10^9$ range.
In many applications of interest, some of the quark propagators are closed --- i.e. they start and end at the same spacetime position --- leading to the problem of computing the
diagonal (in spacetime components) of $D^{-1}$.
This occurs e.g. in the computation of flavor singlet hadron masses such as $m_\eta$ or $m_{\eta^{\prime}}$; in the study of multi-hadron states; in the computation of the strangeness content of the nucleon or the pion-nucleon-sigma term $\sigma_{\pi N}$; or in the study of weak decays of flavored mesons, that we will specifically address in this work.
The construction of closed propagators by means of simple lattice techniques usually implicates huge statistical noise, since it is impossible to sum over space at the insertion point of the
operator that gives rise to the quark loop.
This brings the need of sophisticated computational techniques.
The inversion of the lattice Dirac operator $D$ proceeds by solving linear systems of the form
\begin{equation}
   D\Phi = \eta,
\label{eq:linsyst}
\end{equation}
where $\eta$ a source vector. In its simplest form, $\eta$ is taken to be a point source,
 i.e.\footnote{For simplicity, color, spin, and flavor indices are suppressed.}
\begin{equation}
  \eta(x^\prime) = \delta_{x^{\prime}y}\,.
\end{equation}
This implies that the solution of eq.\,(\ref{eq:linsyst}) yields a ``one-to-all'' propagator: the quark propagator $D^{-1}(x,y) \equiv S(x,y)$ from a single point~$y$ to any other point~$x$ of the lattice, which
corresponds to just one column (or row) of the full propagator matrix. 
In typical simulations of LQCD, solving eq.\,(\ref{eq:linsyst})
to machine precision for all source positions is beyond the
capabilities of even the most powerful supercomputers due to the large dimension of $D$. Therefore, the computation of the full propagator matrix or the inverse diagonal poses a huge computational challenge.
Stochastic volume sources (SVS)~\cite{Bernardson,Dong_Liu} have been long used as a means to access the full propagator matrix by replacing
it with a stochastic estimate such that the entries for all lattice points are available.
These stochastic ``all-to-all'' propagators have been successfully applied in
a number of different
contexts (see e.g. Refs.~\cite{RomeII,Peisa,Foster,SESAM,Cais,Foley,McNeile,Simula,andreas_article,ETM,ETMeta}),
but usually a large computational effort and an appropriately chosen dilution scheme are required in order to sufficiently reduce the intrinsic stochastic noise.
In this work we apply a recently proposed ``probing'' algorithm~\cite{Saad} to the problem of approximating the diagonal entries of a matrix inverse to LQCD computations, and compare it to the use of SVS. In the probing method the linear system of  eq.\,(\ref{eq:linsyst}) is solved for specially designed probing vectors $\eta$ which are constructed by coloring the graph associated with the Dirac operator $D$, and which exploit the sparsity pattern of the inverse matrix.\footnote{An alternative algorithm, based on domain decomposition techniques, has been proposed by the same authors in Ref.~\cite{DDSaad}.}
In addition, we incorporate probing and SVS into the framework of low-mode averaging (LMA)~\cite{LMA_DeGrand,LMA_hartmut}. In the latter the Dirac propagator $S$ is split into a
contribution $S_{}^{\rm l}$ composed of the $N_{\rm low}$ lowest-lying eigenmodes, which are treated exactly, and its orthogonal complement $S_{}^{\rm h}$. In the original formulation of LMA, the contribution $S_{}^{\rm h}$ is computed by means of point sources, whereas here we apply the two aforementioned all-to-all propagator techniques for this task.
Our main target application for this setup lies within a long-term project to study the role of the charm
quark in the $\Delta I = 1/2$ rule for non-leptonic kaon decay~\cite{Hartmut,prl,zm}.
More specifically, we will consider the severe signal-to-noise ratio arising in the computation
of ``eye diagrams'' (penguin contractions), that appear as contributions to $K\to\pi$ transition
amplitudes mediated by the $\Delta S=1$ weak effective Hamiltonian.
In this context, both probing and SVS can be applied to the closed quark propagators appearing
in eye diagrams, which are the main source of statistical noise. SVS will also be applied to propagators
connecting four-fermion operators and interpolating meson operators, which are also connected
to large statistical fluctuations.
Physics applications are discussed in a companion paper~\cite{physpaper};
preliminary results of our study were also presented in Ref.~\cite{proceedings}. 
%


The numerical work we present is carried out on quenched ensembles, and small or only moderately
large lattices. The latest are however suitable enough for our present physics goals,
as discussed in~\cite{physpaper}, while the smallest lattices will be mostly used for
checking technical issues only indirectly related to physics. It has to be stressed that
immediate applications of these techniques are intended in an environment where overlap
fermions are employed; the numerically expensive character of the latter would make the
use of larger lattices rapidly forbidding. Direct comparison with recent works similarly
addressing all-to-all techniques (see e.g. \cite{Blum:2012uh,new_probing,Bali:2009hu,Morningstar:2011ka,Alexandrou:2013wca})
are not straightforward.

The use of quenched configurations deserves some specific remarks. Using them is again
sufficient for the purpose of our immediate intended applications~\cite{physpaper},
and avoids the need to set up a mixed-action approach in case an overlap valence sector
is used in combination with dynamical configurations obtained e.g. using Wilson sea fermions.
It is however worth noting that, based on our knowledge of the Dirac spectrum for
low values of quark masses, quenched ensembles are expected to be worst-case scenarios
for some of the variance issues we are addressing. In particular, the width of the
gap in the Dirac spectrum for computations in the $\epsilon$-regime (see~\cite{physpaper}
for details) is expected to be controlled by the parameter $N_{\rm f}+|\nu|$, where $N_{\rm f}$ is
the number of light dynamical flavours and $\nu$ is the topological charge of the
configuration.\footnote{See e.g.~\cite{bump,Bernardoni:2010nf} and references therein.}
The case $N_{\rm f}=0$ is thus the one where the overlap Dirac operator is expected to be
worse-conditioned. Furthermore, it is also the case where wavefunction properties
may have a largest impact on the variance of fermionic observables~\cite{bump}.
We thus expect that qualitative observations on quenched ensembles will hold for
more realistic simulations where $N_{\rm f}=2(+1)(+1)$ dynamical flavours are employed.


%
The outline of this paper is as follows: in section~\ref{sct:prop_tech} we
review concepts of all-to-all propagator techniques. This comprises stochastic volume sources, the novel probing algorithm and low-mode averaging. In section~\ref{sct:applications} we apply probing and SVS to the computation of traces of quark loop propagators (section~\ref{sct:2pt}) which are the usual building blocks of e.g. flavor singlet two-point functions. Moreover, we combine both SVS and probing with low-mode averaging (section~\ref{sct:eye}), and discuss its application to the study of the $\Delta I = 1/2$ rule.
Section~\ref{sct:conclusion} contains a summary of our findings as well as some concluding remarks.\footnote{During the preparation of this work, a related study appeared in Ref.~\cite{new_probing}, where probing techniques are also applied within the context of LQCD, and a purposely adapted variation dubbed ``hierarchical probing'' has been proposed.}
\section{All-to-all propagator techniques}\label{sct:prop_tech}
For a given gauge field, the propagator $S(x,y)$ from lattice site $y$
to $x$ is obtained as the solution of the linear system
\begin{equation}
\label{eqn:source_method}
 \sum_{z}\sum_{c,\gamma}D_{\alpha\gamma}^{ac}(x,z)
        S_{\gamma\beta}^{cb}(z,y)= 
        \delta_{\alpha\beta}\delta^{ab}\delta_{xy}\,,
\end{equation}
where Latin indices are used to label color and Greek letters denote
spinor components.
Using point sources amounts to solving the linear system for one
particular choice of~$y$. Altogether 12 inversions
must be performed for a single spacetime point: one for each of the $4\times 3$ spin-color combinations. This yields a single column (or row) of the propagator matrix.  Thanks to translational invariance a large class of correlation functions can be defined in terms of these one-to-all propagators.
To illustrate the appearance of closed propagators with a simple example,
consider the generic structure for a two-point function of quark bilinear operators projected to zero spatial momentum, viz.
\begin{equation}
C^{ab}(t)=
\sum_{\vec{x},\vec{y}}\left\langle\bar\psi(x)T^a\Gamma\psi(x)\,\bar\psi(y)T^b\Gamma'\psi(y)\right\rangle,\label{eq:full_2pt_fct_def}
\end{equation}
where $t=x_{0}^{} - y_{0}^{}$, $T^{a,b}$ are ${\rm U}(N_{\rm f})$ flavor generators, $\Gamma^{(')}$ are spin matrices, and spin, flavor, and color indices are contracted inside bilinears.\footnote{In eqs.~(\ref{eq:full_2pt_fct_def},\ref{eq:full_2pt_fct}) quark fields and propagators carry flavor indices.}
After performing the Wick contractions this can be written as
\begin{align}
\nonumber
C^{ab}(t)=\sum_{\vec{x},\vec{y}}&\Big\{ 
 \left\langle\text{tr}\left\lbrace S(x,x)\Gamma T^a\right\rbrace \text{tr}\left\lbrace S(y,y)\Gamma' T^b\right\rbrace \right\rangle_{\rm G}\\
 &\quad -\,\left\langle \text{tr}\left\lbrace\gamma_{5}^{} S_{}(x,y)^{\dagger}\gamma_{5}^{}\Gamma T^a S(x,y)\Gamma'T^b\right\rbrace\right\rangle_{\rm G}\Big\},\label{eq:full_2pt_fct}
\end{align}
where traces are taken over spin, color, and flavor indices, and $\langle\rangle_{\rm G}$
indicates average over gauge configurations taken with the effective action including
the determinant of the massive Dirac operator.
The second term is usually referred to as ``connected'' contribution, while the first term, that survives only when $T^a=T^b=\mathbf{1}$, is called ``disconnected'' contribution. Thus, whenever the disconnected contribution is required (as e.g. in the study of the properties of $\eta'$ mesons), all spacetime-diagonal entries of the propagator matrix are needed. In the following several all-to-all propagator techniques are reviewed which allow, in particular, to tackle the computation of closed propagators.
%

\subsection{Stochastic volume sources}
In the stochastic approach, an ensemble of $N_{\rm r}$ random vectors,
$\left\lbrace \eta^{(r)}(x_0,\vec{x})|r=1,\dots,N_{\rm r}\right\rbrace$,
is generated for each gauge configuration. These  $N_{\rm r}$ ``hits'' 
are created by assigning independent random numbers to all components of the source vector,
i.e. to all lattice sites, color and Dirac indices. Each random
number is drawn from a distribution which is symmetric
about zero in the hit limit ${N_{\rm r}\rightarrow \infty}$, i.e.
\begin{equation}\label{symmetric_about_zero}
  \left\langle \eta^{a}_{\alpha}(x_0,\vec{x}) \right\rangle_{\rm{src}}
  \equiv \lim_{N_{\rm r}\to\infty} \frac{1}{N_{\rm r}}\sum_{r=1}^{N_{\rm r}}
  \big(\eta^{(r)}\big)_{\alpha}^{a}(x_{0},\vec{x})=0. 
\end{equation}
In addition, the sources have to satisfy the orthonormality condition
\begin{equation}\label{orthonormality_condition}
  \left\langle \eta_{\alpha}^{a}(\vec{x},x_{0})
  (\eta^{\dag})_{\beta}^{b}(\vec{y},y_{0})
  \right\rangle_{\rm{src}} = 
  \delta_{x_{0}y_{0}}\delta_{\vec{x}\vec{y}}
  \delta_{\alpha\beta}\delta^{ab}.
\end{equation}
Solving the linear system of eq.~(\ref{eq:linsyst}) for each of the
$N_{\rm r}$ source vectors yields a set of solution vectors
\begin{equation}
  \big(\Phi^{(r)}\big) _{\alpha}^{a}(x)= \sum_{z}\sum_{c,\gamma}
  S_{\alpha\gamma}^{ac}(x,z) \big(\eta^{(r)}\big)_{\gamma}^{c}(z).
\end{equation}
Eventually, the estimate of the entire propagator is defined as the stochastic
average (``hit average'') over the product between solution and random source vectors
\begin{align}\label{estimate}
  \left\langle \Phi_{\alpha}^{a}(x)(\eta^{\dag})_{\beta}^{b}(y)
  \right\rangle_{\rm{src}}
 &= \sum_{z}\sum_{c,\gamma}
    S_{\alpha\gamma}^{ac}(x,z)\,\delta_{zy} \delta_{\gamma\beta}\delta^{cb}\nonumber\\
 &= S_{\alpha\beta}^{ab}(x,y).
\end{align}
In general, elements of $\mathbb{Z}_N$ are very effective~\cite{Bernardson,Dong_Liu} in realizing the condition of eq.\,(\ref{orthonormality_condition}). In this work we follow Foster
and Michael\,\cite{Foster} and draw separate elements of
$\mathbb{Z}_2$ for the real and imaginary parts of the source vector,
i.e. 
\begin{equation}
\left(\eta^{(r)}\right)_{\alpha}^{a}(x)\in   \mathbb{Z}_2\otimes \mathbb{Z}_2=\left\lbrace
   \frac{1}{\sqrt{2}}\left(\pm1 \pm  i \right)  \right\rbrace.
\end{equation}
Experience shows that a random source vector, which is distributed
over the entire spacetime lattice, usually leads to a poor
signal for hadronic correlation functions. An essential step towards a
significant reduction of the introduced intrinsic stochastic noise is taken by restricting the support of the source vector to individual timeslices, spacetime points, Dirac or color components. This technique is referred to as dilution~\cite{Foley}, and amounts to breaking up
a single vector into $N_{\rm d}^{}$ vectors, each consisting of fewer non-zero random numbers $Z_{i}$.
It can be illustrated as follows
\begin{align}
 \bordermatrix{%
  & \eta\cr
  & Z_{1}^{}\cr
  & Z_{2}^{}\cr
  & Z_{3}^{}\cr
  & \vdots\cr
  & Z_{N_{\rm d}^{}}\cr}
\longrightarrow
\bordermatrix{%
  & \eta_{1}^{}\cr
  & Z_{1}^{}\cr
  & 0\cr
  & 0\cr
  & \vdots\cr
  & 0\cr}
\bordermatrix{%
  & \eta_{2}^{}\cr
  & 0\cr
  & Z_{2}^{}\cr
  & 0\cr
  & \vdots\cr
  & 0\cr}\bordermatrix{%
  & \eta_{3}^{}\cr
  & 0\cr
  & 0\cr
  & Z_{3}^{}\cr
  & \vdots\cr
  & 0\cr}\dots\bordermatrix{%
  & \eta_{N_{\rm d}^{}}^{}\cr
  & 0\cr
  & 0\cr
  & 0\cr
  & \vdots\cr
  & Z_{N_{\rm d}^{}}\cr
 }.\nonumber
\end{align}
Thus, the variance of the estimated inverse matrix can be decreased by increasing either
$N_{\rm r}^{}$ or $N_{\rm d}^{}$, i.e. by enlarging the number of hits or by a higher degree of dilution
respectively. In many applications (see e.g. Refs.~\cite{Foley,SVS_BULAVA}) it has been found that dilution
outperforms the application of multiple hits. However, for a fixed number of inversions the approach and dilution scheme leading to an optimal variance reduction has to be determined for the correlation function at hand~\cite{SVS_Wilcox}.
In particular, time-dilution is widely used in the computation of hadronic properties\,\cite{Foster,Cais,Foley}. 
In this scheme the non-zero components of random source vectors are restricted to single timeslices, i.e.
\begin{equation}
\eta(\vec{x},t)=\sum_{j=0}^{N_{t}^{}-1}\eta_{j}^{}(\vec{x},t),\quad \eta_{j}^{}(\vec{x},t)=0 \quad\text{if } t\neq j.
\end{equation}
For the computation of connected diagrams (see section~\ref{sct:applications}), time dilution is an indispensable tool for obtaining well behaved signals~\cite{Foley}.

\subsection{Probing}\label{sct:PROBING}
%
The probing method explored here has been developed by Tang and Saad~\cite{Saad}, and approximates the diagonal of the inverse matrix $S$ denoted by $\mathcal{D}(D^{-1})=\mathcal{D}(S)$. The method is designed for sparse matrices whose inverse has a strong decaying behavior --- that is, the values of the off-diagonal inverse matrix entries are required to fall off rapidly when moving away from the main diagonal. The target diagonal is approximated or ``probed'' by means of matrix-vector multiplications. The associated so-called probing vectors are generated by a procedure of graph coloring. A form of graph coloring has been originally
exploited to evaluate sparse Jacobian and Hessian matrices~\cite{GRAPH_COLORING_3,GRAPH_COLORING_2,GRAPH_COLORING_1,GRAPH_COLORING_4}.
Unlike first attempts based on random number vectors~\cite{Bekas}, the kind of probing vectors
considered in this algorithm is able to exploit the sparsity pattern of $S$.
Given a set of $s$ probing vectors $v_{i}^{}$, i.e. $V_{s}^{}:=\:\{v_{1}^{},v_{2}^{},\ldots,v_{s}^{}\}\in\mathbbm{R}^{N\times \rm s}, s\in\mathbbm{N}$, the diagonal of the inverse of the $N\times N$ dimensional Dirac operator can be approximated as~\cite{Saad}
\begin{equation}
 \mathcal{D}(S)\approx\mathcal{D}(SV_{s}^{}V_{s}^{T})\mathcal{D}_{}^{-1}(V_{s}^{}V_{s}^{T}),
 \label{eq:est_diag}
\end{equation}
where the notation $\mathcal{D}(M)$ refers to a matrix which has the same diagonal as the
square matrix $M$, and zeroes in all off-diagonal entries; and $\mathcal{D}_{}^{-1}(M)$ is
its inverse.
An intuitive understanding of eq.~(\ref{eq:est_diag}) follows by noting that the approximation would be exact in the limit $s=N$ when $V_{N}^{}$ is a unitary matrix, because this implies $V_{s}^{}V_{s}^{T}=V^{T}_{N}V_{N}^{}=\mathbf{1}$. More formally, the method is based on the proposition proven in Ref.~\cite{Bekas}, whereby eq.~(\ref{eq:est_diag}) holds exactly if each $i$-th row of $V_{s}^{}$ is orthogonal to all those rows $j$ of $V_{s}^{}$ for which the $(ij)$-th entry of $S$ is non-vanishing. Whereas orthogonality is required for the rows of $V_{s}^{}$, the columns do not have to satisfy this property.
The probing method amounts to finding a suitable set $V_{s}^{}$ of probing vectors $v_{i}^{}$,
with $s\ll N$, that effectively recovers the spacetime-diagonal entries of the matrix $S$. To this end,
the $s$ columns of $V_{s}^{}$ have to be constructed in such a way that $\mathcal{D}(S)$ is minimally
contaminated by contributions from off-diagonal elements of $S$ via the matrix-vector multiplication
$SV_{s}^{}$. While contributions from elements far away from the diagonal are expected to be small due to the
assumed decay law, contributions to $S(x,x)$ from sites close to $x$ will be non-negligible,
and have to be suppressed by properly choosing the zero entries in the probing vectors.
For instance, having a probing vector $v=[1~0\dots 0~1~0\dots 0~1~0 \dots 0]^{T}$, the $i$-th '1' is responsible for retrieving the $i$-th element of the diagonal of the inverse, while the other non-zero
entries will pick up off-diagonal contributions, of which only small ones should be kept.
One thus has to find an efficient way of obtaining maximal separations between non-zero entries
in terms of lattice distance.
The actual procedure to determine the probing vectors is based on standard graph theory (see e.g. Ref.~\cite{SAAD_BIBLE}).
Assuming that elements of $S$ decay in entries which are far away from the positions of the non-zero elements of $D$, the small entries of $S$ can be dropped. This leaves a sparse matrix $S_{\epsilon}^{}$. In the context of probing this corresponds to dropping the elements of $S(i,j)$ whose vertices $i$ and $j$ are farther apart than the distance $p$ in the graph of $D$. The distance $p$ is defined as the number of links connecting the two vertices $i$ and $j$. 
Then, the sparsity pattern of $S_{\epsilon}^{}$ can be approximated by the sparsity pattern of some power $p$ of $D$. For a strong decay behavior this yields $S_{\epsilon}^{} \approx S$ for small $p$.
In other words, since the propagator (or, more in general, the inverse) has some decay property, it can be approximated by a matrix that has strictly zero components when the distance to the diagonal is beyond a certain threshold. The inverse of such a matrix would be some approximate Dirac operator (the form of which depends on how many lines of non-zero components parallel to the inverse diagonal are allowed), and the resulting structure provides a choice for the probing vectors by constructing them such that the procedure would be exact if $S_\epsilon^{}$ would be exact.\footnote{For a deeper understanding and discussion of relating the sparsity pattern with the notion of off-diagonal decay, see the original paper of Ref.~\cite{Saad} and references therein.}
The starting point in constructing the set of probing vectors is to assign colors to the adjacency graph associated with $S_{\epsilon}^{}$ by means of the Greedy Multicoloring Algorithm (see e.g. Ref.~\cite{SAAD_BIBLE}).
The latter requires that no adjacent ($p=1$), next-to adjacent ($p=2$), next-to-next adjacent ($p=3$), etc. vertices have the same color. This is summarized in Algorithm~\ref{algo:color}.
After having colored the adjacency graph, the probing matrix $V_{s}^{} = \:\{v_{1}^{}, v_{2}^{}, \ldots, v_{s}^{}\}$ is derived from the colored graph according to
\begin{equation}
(V_{s}^{})_{}^{jk}=
\begin{cases}
\begin{array}{ll}
1,\:& \text{if Color}(j)=k,\\
0,\:& \text{otherwise}.
\end{array}\end{cases}
\label{eq:V_s}
\end{equation}
The total number of probing vectors $s$ is equivalent to the number of colors used during the coloring process, i.e. $s={\rm max}\{k\}$. Following the procedure of eq.~(\ref{eq:V_s}), each row of $V_{s}^{}$ will contain a single non-zero entry and the diagonal of $V_{s}^{}V_{s}^{T}$ is filled with ones. Thus, $\mathcal{D}(V_{s}^{}V_{s}^{T})=\mathcal{D}_{}^{-1}(V_{s}^{}V_{s}^{T})=\mathbf{1}$.
\begin{algorithm*}[t!]
\KwData{Adjacency graph corresponding to a $N\times N$ matrix}
\KwResult{Colors assigned to the vertices of the graph}
Initialization:\\
\For{j=1 to $N$}
{ Set \text{Color}(j) = 0\\
 }
 \For{j=1 to  $N$}{
 Set \text{Color}(j) = min\{  $k > 0~|~ k \neq \text{Color}(l)~ \forall l\in \text{Adjacent(j)}$ \}
 }
 \caption{Greedy Multicoloring Algorithm}
 \label{algo:color}
\end{algorithm*}
\begin{figure}[t!]
\begin{center}
 \includegraphics[scale=0.55]{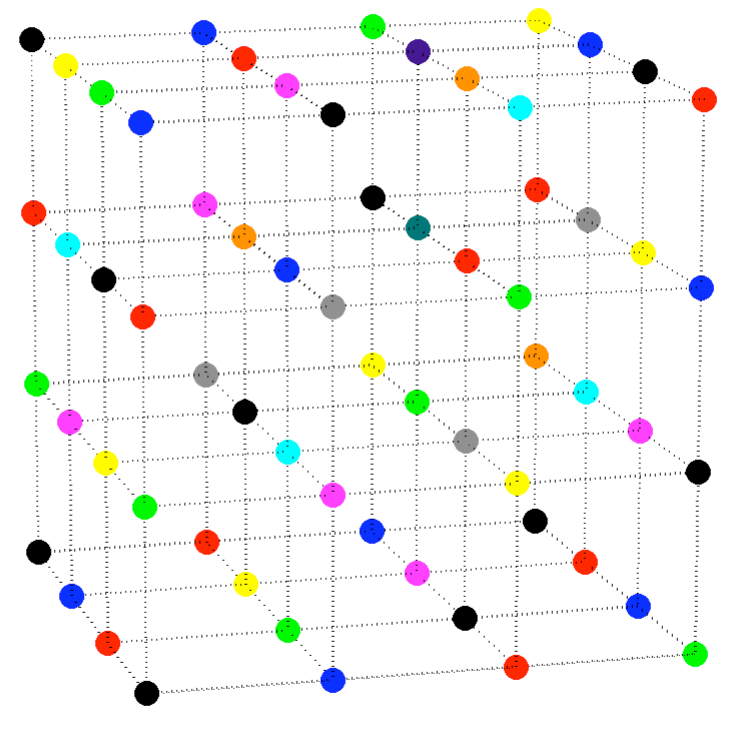}
\end{center}
\caption {Coloring a 3-dimensional mesh for $p=2$ with no boundary conditions. The coloring is done starting in the left, front corner of the bottom plane. Then, moving from front to back, left to right and finally bottom to top.} 
\label{fig:COLORING}
\end{figure}
In order to illustrate the above steps, Figure~\ref{fig:COLORING} shows a three-dimensional example 
where the vertices of the adjacency graph have been colored for four vertices in each direction.
Boundary conditions are neglected in the plot. 
Distance-$p$ coloring proceeds by assuming that vertices connected via $p$ links (dotted lines) differ in their assigned colors. Consequently, for $p=1$ all nearest-neighbors must have different colors, and the total number of required  colors is $s=2$. In Figure~\ref{fig:COLORING} we have set $p=2$ --- i.e. vertices which can be connected via one or two links must have different colors, which leads to $s=11$ colors.  According to the assignment directive of eq.~(\ref{eq:V_s}), the resulting probing matrices with a single non-zero entry in each row are given, respectively, by
\begin{equation}
V_{2}=\left[\begin{array}{cc}
1 & 0\\
0 & 1\\
1 & 0\\
\vdots & \vdots\\
1 & 0\\
0 & 1
\end{array}\right],~V_{11}=\left[\begin{array}{ccccc}
1 & 0 & 0 & \ldots & 0\\
0 & 1 & 0 & \cdots & 0\\
0 & 0 & 1 & \cdots & 0\\
1 & 0 & 0 & \ldots & 0\\
\vdots & \vdots & \vdots & \ldots & \vdots\\
0 & 0 & 0 & \ldots & 1
\end{array}\right].
\end{equation}
The columns of the probing matrix serve as source vectors for solving the usual linear system
\begin{equation}
Dx_{i}^{}=v_{i}^{},\qquad\longrightarrow\qquad x_{i}^{}=D_{}^{-1}v_{i}^{}\label{eq:Axv}.
\end{equation}
The inversions yield a set of $s$ solution vectors $X_{s}^{}:=\:\{x_{1}^{},x_{2}^{},\ldots,x_{s}^{}\}:=S_{\epsilon}^{}V_{s}^{}$.
Having $\mathcal{D}_{}^{-1}(V_{s}^{}V_{s}^{T})=\mathbf{1}$, the expression of eq.~(\ref{eq:est_diag}) implies an approximation of the inverse diagonal of the form
\begin{equation}
\mathcal{D}(S_{\epsilon})\approx\mathcal{D}(X_{s}^{}V_{s}^{T}).
\label{eq:DXV}
\end{equation}
Compared to the exact computation of the inverse diagonal, the probing method significantly reduces the computational effort by solving only ${s}$ linear equations with ${s} \ll N$. The inversions can be done with any standard iterative solver.
When applying the probing method to the lattice Dirac operator, the coloring is performed in space and time coordinates only, i.e. the internal spin and color structure (which is not sparse) is not considered in the coloring procedure. As in the standard point source approach, the non-zero entries of $V_{s}$ represent a unit matrix in spin-color space, i.e. $\mathbf{1}=\mathbf{1}_{12\times 12}$.
Consequently, the computation of $x_{i}^{}$ involves twelve separate inversions --- one for each of the twelve spin-color combinations.
Algorithm~\ref{algo:probe} summarizes the probing algorithm used for approximating the diagonal of the inverse lattice Dirac operator.
The number of colors $s$ needed for a given $p$ is essentially insensitive to the volume, as can be seen in Table~\ref{tab:color}.
%

%
\begin{algorithm*}[t!]
\KwData{Lattice Dirac operator $D$, matrix of dimension $N \times 12$}
\KwResult{Approximation of the inverse diagonal $\mathcal{D}(D^{-1}_\epsilon)=\mathcal{D}(S_{\epsilon}^{})$}
Initialization:\\
\For{any p}
{
Color the vertices of the $N\times N$ adjacency graph in space and time coordinates; no coloring in spin nor color indices. Apply boundary conditions.\\
Construct probing matrix $V_{s}^{}=\:\{v_{1}^{},v_{2}^{},\ldots,v_{s}^{}\}$ following the assignment of eq.~(\ref{eq:V_s}), viz.
$(V_{s}^{})_{}^{jk} =\mathbf{1}_{12\times 12} \: \text{ if Color}(j)=k \text{, or } \: \: 0 \: \text{ otherwise}$\\
 \For{i=1 to  $s$}{
 \For{l=1 to 12}{
  Solve linear system of eq.~(\ref{eq:Axv}) for each spin-color index $l$ using a Krylov subspace method of choice. 
 }
 Construct $x_{i}^{}$\\
 }
 Construct $X_{s}^{}:=\:\{x_{1}^{},x_{2}^{},\ldots,x_{s}^{}\}$ \\
 Compute $\mathcal{D}(D^{-1}_\epsilon) = \mathcal{D}(X_{s}^{}V_{s}^{T})$\\
 Set $\mathcal{D}(S):=\mathcal{D}(S_{\epsilon}^{})=\mathcal{D}(D^{-1}_\epsilon)$
 }
 \caption{Algorithm to probe the diagonal of the Dirac operator}
 \label{algo:probe}
\end{algorithm*}
\begin{table}[t!]
\begin{center}
\begin{tabular}{r |c c c c c}
\noalign{\vskip0.3ex} \hline\hline\noalign{\vskip0.3ex}  & p=1 &
p=2 & p=3 & p=4 & p=5  \\
\noalign{\vskip0.3ex} \hline\noalign{\vskip0.3ex}
   $8^{4}$                   &    2    &    22    &    36     &   117    &     175   \\
   $16^{4}$                 &    2    &    23    &    36     &   121    &     175  \\
   $32$ x $16^{3}$    &    2    &    22    &    37     &   123    &     173  \\
   $48$ x $24^{3}$    &    2    &    23    &    35     &   122    &     176 \\
   $64$ x $32^{3}$    &    2    &    23    &    36     &    120  &       174 \\       
\hline\hline
\end{tabular}
\caption{\small Number of colors required for coloring with distance $p$ for different lattice sizes of the 4-dimensional LQCD setup.}
\label{tab:color}
\end{center}
\end{table}
The probing method, on the other hand, has potential shortcomings.
While increasing the distance $p$ yields a higher precision, the computational cost
increases rapidly with $p$, and soon becomes prohibitively expensive.
Besides, the choice of the appropriate $p$ which ensures a desired accuracy is not known \textsl{a priori}.
In particular, if a certain $p$ does not yield a required accuracy, an increase of $p$ can not reuse previous computations since its probing vectors are not related to the previous ones.
As mentioned in the introduction, during the preparation of this work a promising version of probing, that aims at taking into account more structural details of lattice Dirac operators, has been proposed in Ref.~\cite{new_probing}. This so-called ``hierarchical probing' elegantly allows to reuse the computations of prior choices of $p$, and furthermore hints at smaller variances for fixed cost. This comes at the price
of a significant increase in the complexity of the algorithm, with respect to the one we are considering here.
%
%
\subsection{Low-mode averaging}\label{sct:lma}
%
In the spectral representation the quark propagator can be written as
\begin{equation}\label{eqn:LMA_0}
S(x,y)=\frac{1}{V}\sum_{i=1}^{N}
\frac{v_{i}^{}(x)\otimes v_{i}^{\dag}(y)}{\lambda_{i}+m},
\end{equation}
where $m$ is the quark mass and $\lambda_i$ and $v_i$ are eigenvalues and eigenmodes of the Dirac operator, respectively, viz.
\begin{equation}
Dv_i=\lambda_iv_i\,.
\end{equation}
We will assume the normalization $\sum_x|v_i(x)|^2 = V$, where $V$ is the four-volume.
The structure of the low-lying spectrum of the Dirac operator is determined by the physics of spontaneous symmetry breaking of QCD chiral symmetry. In particular, the spectral density around zero does not vanish in infinite volume; and in a finite volume there are arbitrarily small eigenvalues with typical spacings determined by the value of the chiral condensate. For small values of $m$, modes with very small eigenvalues will have a huge weight in the spectral decomposition of eq.~(\ref{eqn:LMA_0}), which can induce very large statistical fluctuations in correlation functions if their associated wavefunctions are not sampled accurately.
Low-mode averaging (LMA)~\cite{LMA_DeGrand,LMA_hartmut} aims at tackling this problem by treating the $N_{\rm low}$ lowest eigenmodes exactly, while separating them from the higher modes by truncating the sum over eigenmodes.
The quark propagator is thus split into a ``low-part'' $S^{\rm l}$ and the ``high-part'' $S^{\rm h}$, viz.
\begin{align}
S(x,y)&=S^{\rm l}(x,y)+ S^{\rm h}(x,y)\nonumber\\
&=\frac{1}{V}\sum_{i=1}^{N_{\rm low}}
\frac{v_{i}^{}(x)\otimes v_{i}^{\dag}(y)}{\lambda_{i}+m} + S^{\rm h}(x,y).\label{eqn:LMA}
\end{align}
The high-part lives in the orthogonal
complement of the subspace spanned by the $N_{\rm low}$ lowest modes --- inverting eq.~(\ref{eqn:LMA})
\begin{equation}
S^{\rm h}(x,y)=\Big(\mathbf{1}-\frac{1}{V}\sum_{i=1}^{N_{\rm low}}v_{i}^{}(x)\otimes v_{i}^{\dag}(y)\Big)S(x,y).
\label{eq:projection}
\end{equation}
Splitting the quark propagator into high- and low-mode propagators results in a formal splitting of the correlation function as well --- for instance, the two-point function in eq.~(\ref{eq:full_2pt_fct}) can be immediately seen to decompose in contributions $C^{\rm ll},C^{\rm hl},C^{\rm hh}$, where superscripts denote the number of low-mode and high-mode propagators, respectively.
Albeit not being physical observables, the individual terms can be treated independently in order to improve their statistical signal (taking advantage of symmetries, increasing $N_{\rm low}$ etc.). For instance, translational invariance can be exploited to average over the spacetime entries of the low-modes, which are known exactly, whenever two instances of $S^{\rm l}$ meet at the same location.
In a straightforward version of LMA, the number $N_{\rm low}$ of modes that are treated exactly can be increased until the signal-to-noise ratio in the relevant correlation functions is brought under control. In this setup, the contribution $C^{\rm ll}$ to a two-point function will be known at all spacetime points, but the contribution $C^{\rm hl}$ (and, obviously, also $C^{\rm hh}$) will be known for a fixed position of one of the operators only. Alternatively, a relatively low value of $N_{\rm low}$ can be taken, after which low modes are used as sources to compute extended propagators.  More precisely, by taking $\Gamma v_{i}$ as the source
(where $\Gamma$ is some spin matrix specifying the desired two-fermion operator insertion),
and by restricting the vector to the timeslice $t=t_{f}$, the solution of the inversion is
\begin{equation}
S_{i}^{\rm ext,\Gamma}(x;y_{0})=\left(\frac{a}{L}\right)^{3}\sum_{\vec{y}}S^{\rm h}(x,\vec{y};t_{f})\Gamma v_{i}^{}(\vec{y};t_{f})\label{constructionI}.
\end{equation}
An implicit average over the spacial coordinate $\vec{y}$ is thus automatically performed.
This strategy has been found to be advantageous in terms of computational cost at fixed
signal-to-noise ratio for various two- and three-point functions in Refs.~\cite{LMA_DeGrand,LMA_hartmut}.

\subsection{Hybrid approach}\label{sct:HYBRID}
%
%
A hybrid approach as proposed e.g. in Ref.~\cite{Foley} is the straightforward combination of LMA with other all-to-all propagator techniques.
In the original LMA formulation, the high-part $S_{}^{\rm h}$ of the decomposed propagator is computed by means of standard point source inversions --- or, where possible, put into extended propagators, as explained
above. Here, all-to-all techniques in form of stochastic volume sources or probing are applied to compute the orthogonal complement, such that, in particular, the computation of closed quark loops at all spacetime points can be tackled. 
In the next section we will explore this hybrid strategy for two- and three-point functions relevant for light-meson physics, using SVS and probing for the estimation of $S^{\rm h}$.

\section{Applications}\label{sct:applications}
\subsection{Closed propagators}\label{sct:2pt}

\begin{figure*}[t!]

	\hspace*{-3mm}
    \begin{minipage}{.5\linewidth}
      \begin{center}
      \includegraphics[scale=0.32]{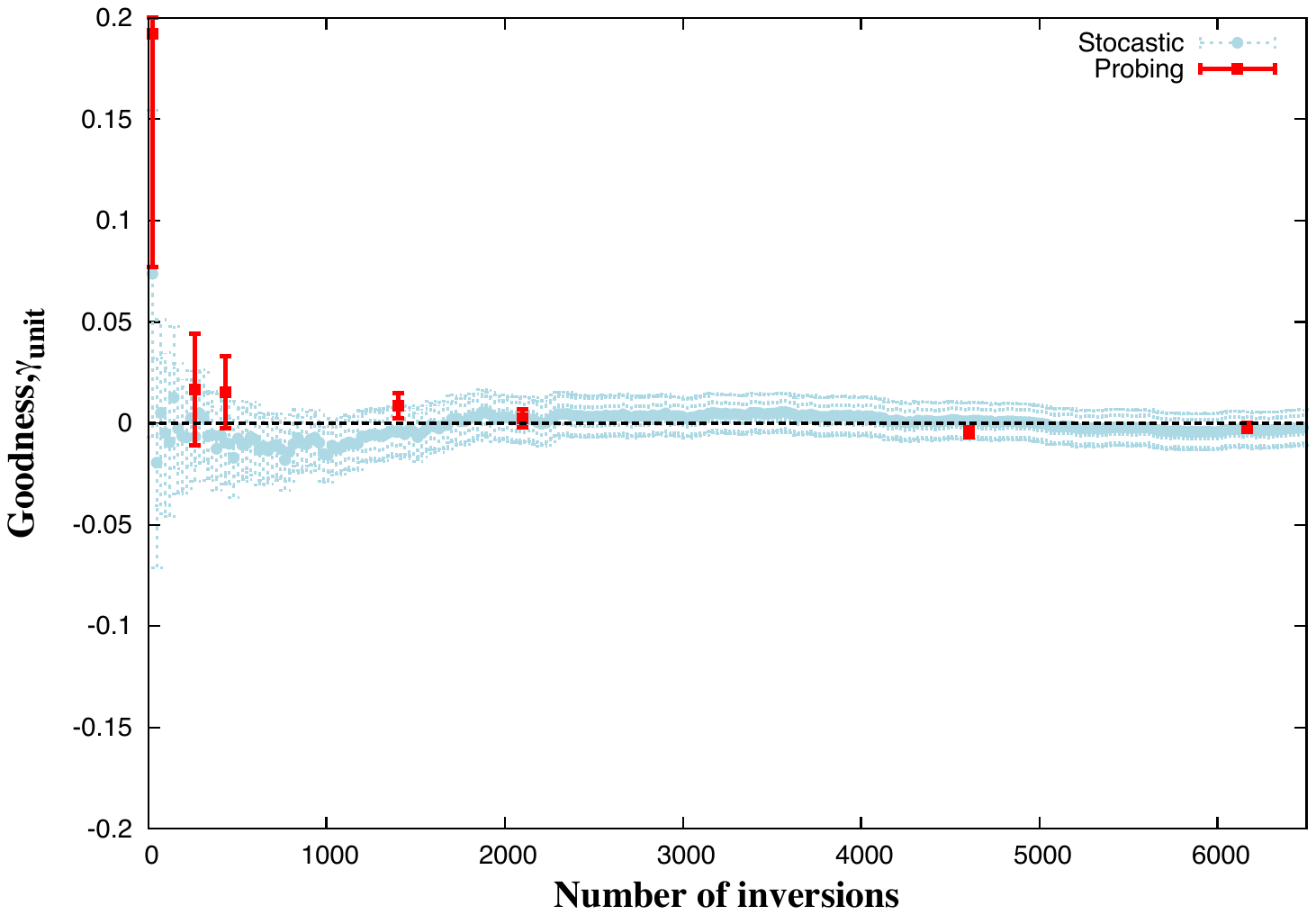}
      \end{center}
    \end{minipage}
    \begin{minipage}{.5\linewidth}
      \begin{center}
      \includegraphics[scale=0.32]{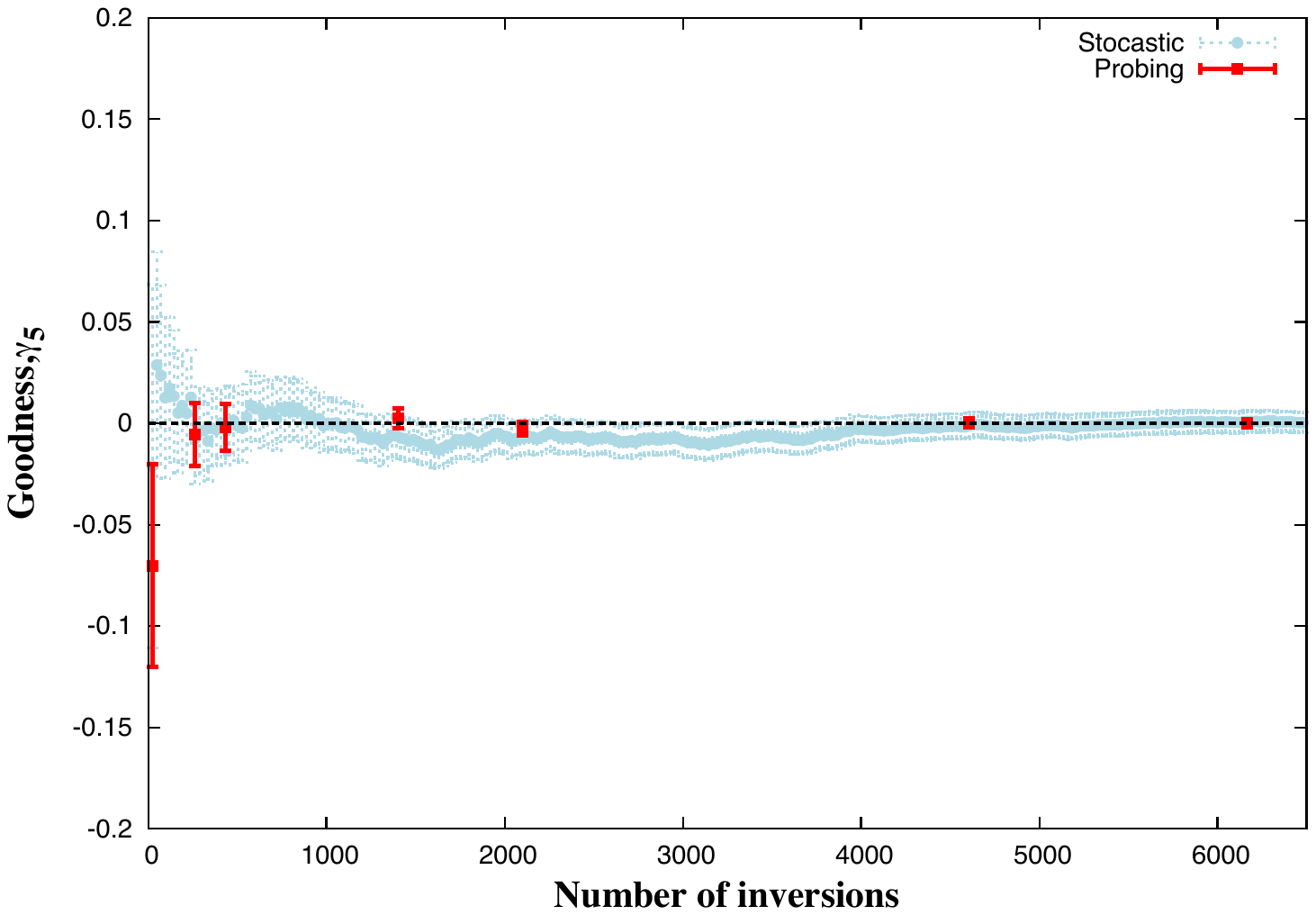}
      \end{center}
    \end{minipage}
    
	\hspace*{-3mm}
    \begin{minipage}{.5\linewidth}
      \begin{center}
      \includegraphics[scale=0.32]{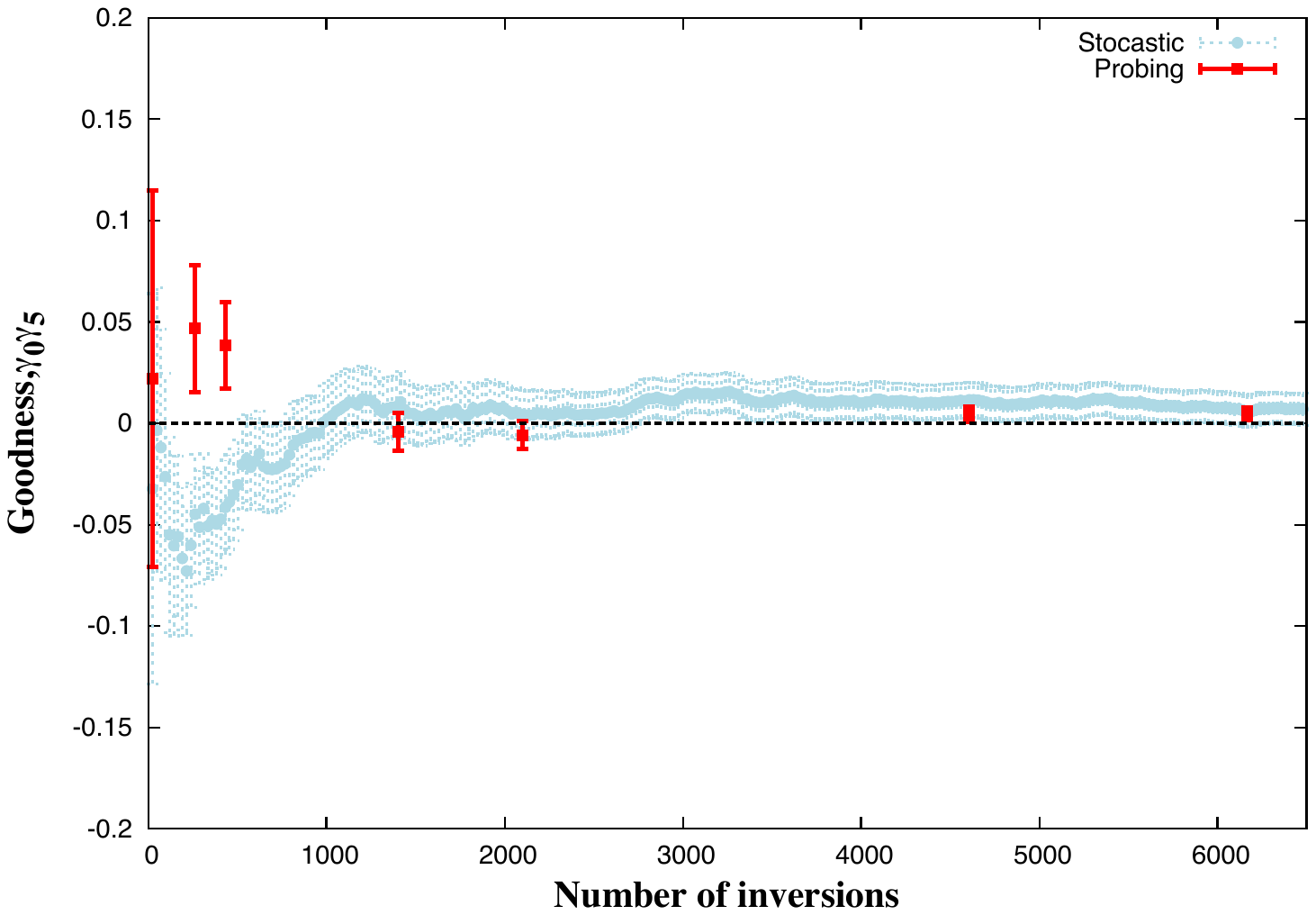}
      \end{center}
    \end{minipage}
    \begin{minipage}{.5\linewidth}
      \begin{center}
      \includegraphics[scale=0.32]{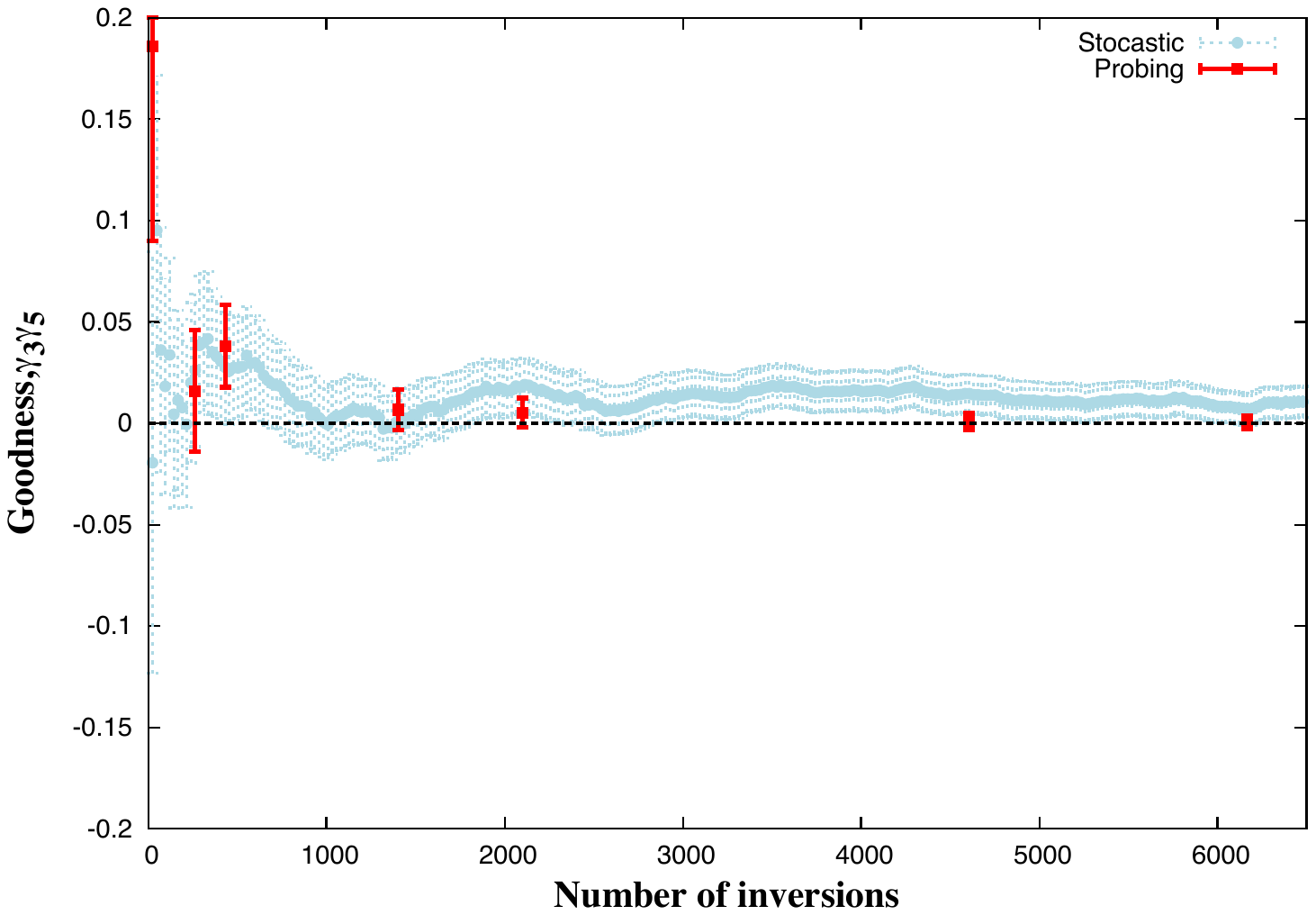}
      \end{center}
    \end{minipage}

\caption{Value of the quantity $\Delta_\Gamma$ in eq.~(\ref{eq:GOODNESS}) for $\Gamma=\mathbf{1},\gamma_5,\gamma_0\gamma_5,\gamma_3\gamma_5$ (in clockwise order starting from the left upper corner).
Red (solid) bars correspond to the probing method for $p=2,\dots,7$, whereas the results of the stochastic technique are denoted by blue (dashed) bars.
A value of zero  corresponds to the exact solution.
}
\label{fig:GOODNESS_05}
\end{figure*}

%
In this section we will focus on exploring the accuracy of SVS and probing methods for computing  closed quark propagators. To that purpose, we will consider expectation values of quark bilinears,
\begin{equation}\label{eqn:loop}
-\langle\bar\psi\Gamma\psi\rangle = \sum_{x}\left\langle\text{tr}\left\lbrace S(x,x)\Gamma\right\rbrace \right\rangle \equiv {\rm Tr}_\Gamma[S(x,x)].
\end{equation}
Due to the spacetime symmetries of the regularized theory,
all expectation values have to vanish after the average over gauge configurations is taken, except in the case $\Gamma=\mathbf{1}$, which provides the bare quark condensate.
The amplitude of the fluctuations around zero will provide a measure of sampling noise.
We have performed our computations on dynamical gauge configurations with $N_{\text{f}}^{} = 2$
flavors of dynamical $O(a)$-improved Wilson fermions, generated using the deflation-accelerated DD-HMC algorithm~\cite{DDHMC}.\footnote{The code is available at \url{http://luscher.web.cern.ch/luscher/DD-HMC/index.html}.} The latter combines domain-decomposition (DD) methods~\cite{SAAD_BIBLE} with the Hybrid Monte Carlo (HMC) algorithm~\cite{HMC} and the Sexton-Weingarten multiple-time integration scheme~\cite{Sexton}. The inversions of the Wilson-Dirac operator are performed using a Schwarz-preconditioned generalized conjugate residual (SAP+GCR) algorithm~\cite{SAP_GCR}.
Two lattices sizes are considered, at fixed lattice spacing given by $\beta=5.30$: an $8^{4}$-lattice (with light quark mass given by the hopping parameter $\kappa=0.13625$) and a $16^{4}$-lattice ($\kappa=0.13620$).
The ensembles consist of 150 and 100 thermalized configurations, respectively; in both cases
the configurations are separated by 100 HMC trajectories of length $\tau=1/2$.
Volume-filling stochastic sources are diluted in space (even-odd), spin and color but not in time.
We will quote results for $\Gamma=\mathbf{1},\gamma_{5}^{},\gamma_{0}^{}\gamma_{5}^{},\gamma_{3}^{}\gamma_{5}^{}$.
In order to measure the quality of the approximation to $S(x,x)$ with respect to the exact solution, we introduce the quantity
\begin{equation}
\Delta_\Gamma \equiv {\rm Tr}_\Gamma[S(x,x)]_{\text{exact}} - {\rm Tr}_\Gamma[S(x,x)]_{\text{method}}.\label{eq:GOODNESS}
\end{equation}
While this observable is too expensive for $16^4$ lattices, we have been able to measure it in our $8^4$ lattices by consecutively putting point sources on all spacetime locations of the lattice. Here ``method'' refers either to the stochastic volume source technique or the probing approach.
The closer $\Delta_\Gamma$ gets to zero, the better the exact result will be approximated.
Results for $\Delta_\Gamma$ as a function of the computational cost are shown in Figure~\ref{fig:GOODNESS_05}. 

\begin{figure*}[t!]

	\vspace{-3mm}
	\begin{center}
    \includegraphics[scale=1.6]{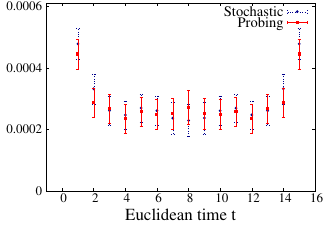}
	\end{center}
	\vspace{-5mm}

\caption{Disconnected contribution to the two-point function of the
singlet axial current in our $16^4$ lattice. Results from probing with $p=4$ and SVS with
the same computational cost are compared.}
\label{fig:2p_singlet}
\end{figure*}

Our findings can be summarized by the following observations: 
\begin{itemize}
\item For both techniques, the approximation of the exact solution has a comparable
accuracy in all cases.
In the case $\Gamma=\mathbf{1}$ (the only non-vanishing case after gauge average), one
can furthermore study the deviation from the expectation value, and see that it is
very small. This is likely due to the fact that the bare 
expectation value $\langle\bar\psi\psi\rangle$ is completely dominated by an ultraviolet divergence inversely proportional to the third power of the lattice spacing, which is essentially insensitive to the fluctuations that affect signal-to-noise ratios in other quantities, and are mostly encoded in off-diagonal spin entries. 
\item For $p=2,3$ probing and SVS yield comparable errors at comparable cost, whereas as soon as $p\geq4$ the deviation from the exact results is considerably smaller for the probing method. Furthermore, the errors of the SVS procedure tend to saturate for numbers of inversions at which an increase of $p$ still improves the accuracy of probing.
\item A choice of $p\geq6$ approximates the exact solution to a degree of accuracy much higher than
the one provided by SVS. However, the computational cost involved might be prohibitive in many practical applications. The difference between the two methods is quite dependent on the operator insertion
considered; observable-specific studies will thus be needed, in general.
\end{itemize}
The plots illustrate as well the shortcoming of having little freedom in tuning the cost of the probing procedure, exemplified by the large increase of computational effort in the steps $p=3\to p=4$ or $p=5\to p=6$.
\begin{figure*}[t!]

	\hspace*{-8mm}
    \begin{minipage}{.5\linewidth}
      \begin{center}
      \includegraphics[scale=0.32]{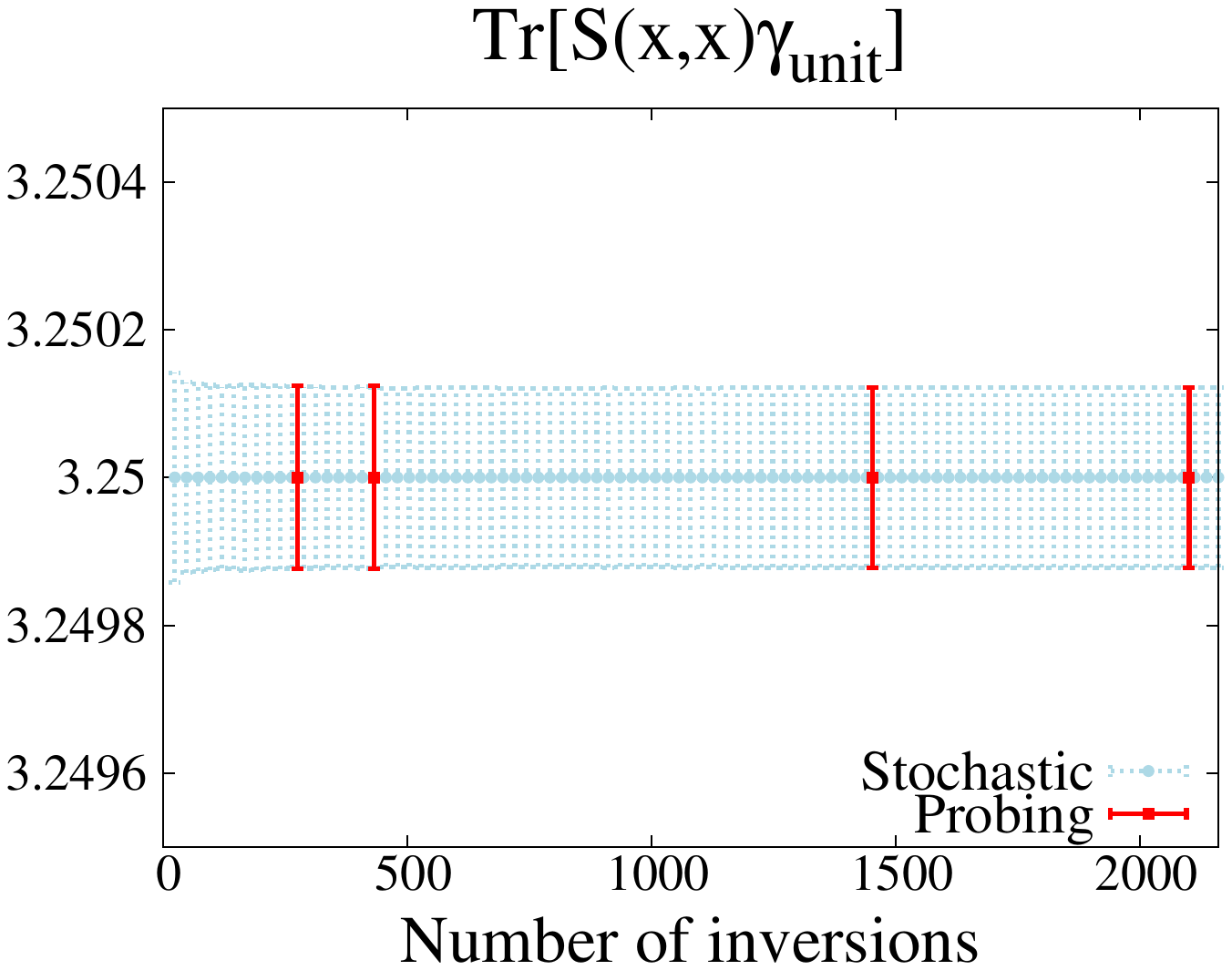}
      \end{center}
    \end{minipage}
    \begin{minipage}{.5\linewidth}
      \begin{center}
      \includegraphics[scale=0.32]{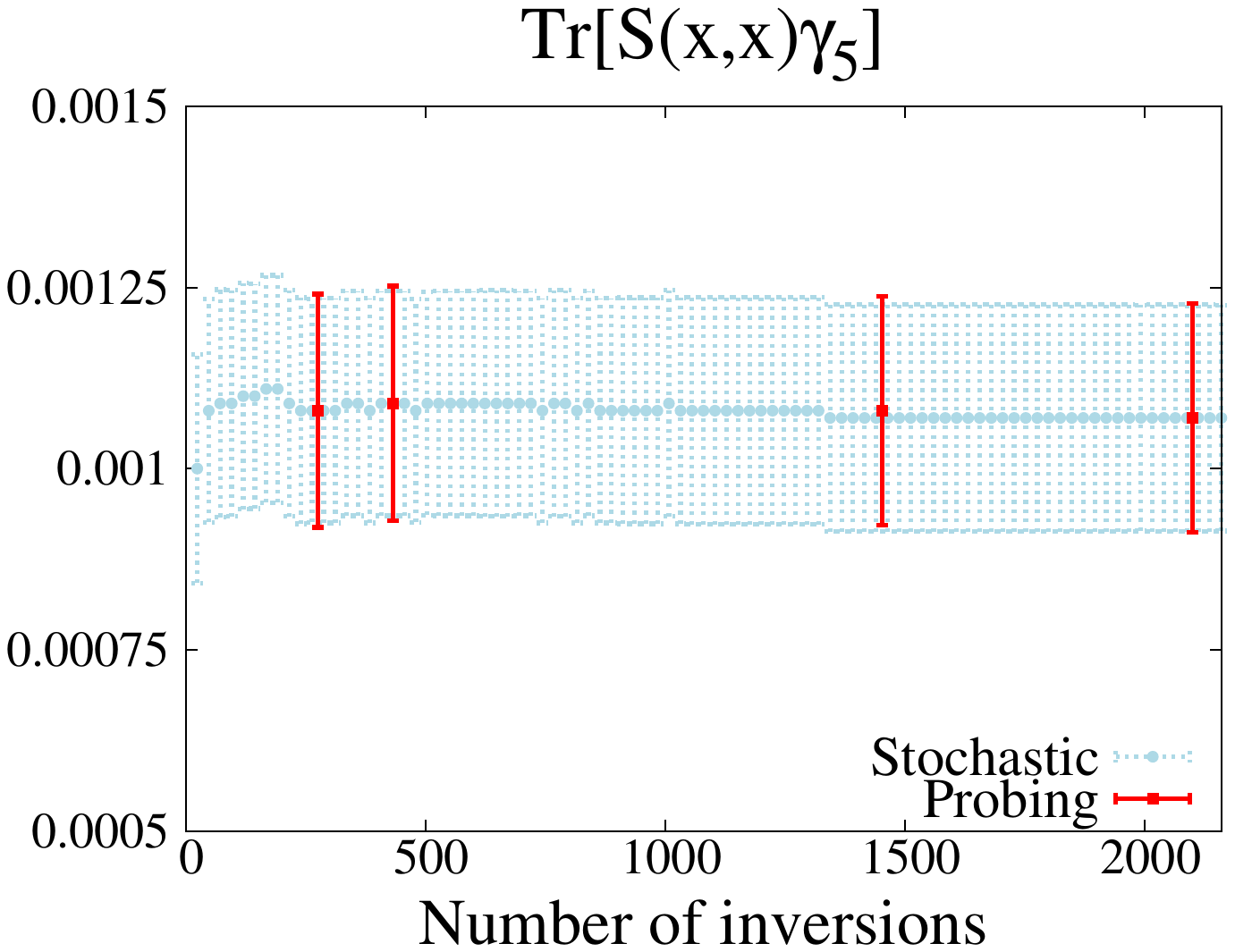}
      \end{center}
    \end{minipage}
    
	\hspace*{-8mm}
    \begin{minipage}{.5\linewidth}
      \begin{center}
      \includegraphics[scale=0.32]{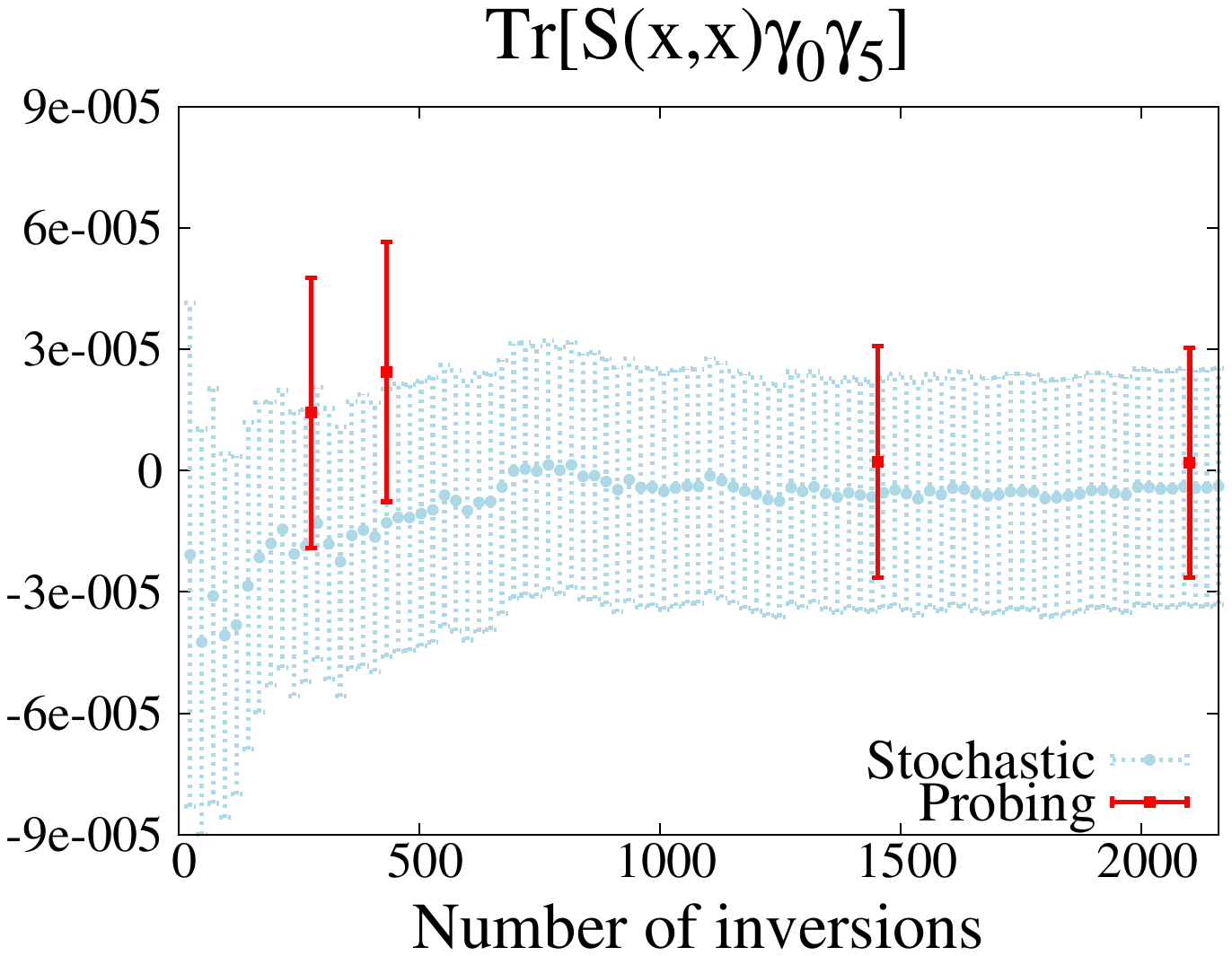}
      \end{center}
    \end{minipage}
    \begin{minipage}{.5\linewidth}
      \begin{center}
      \includegraphics[scale=0.32]{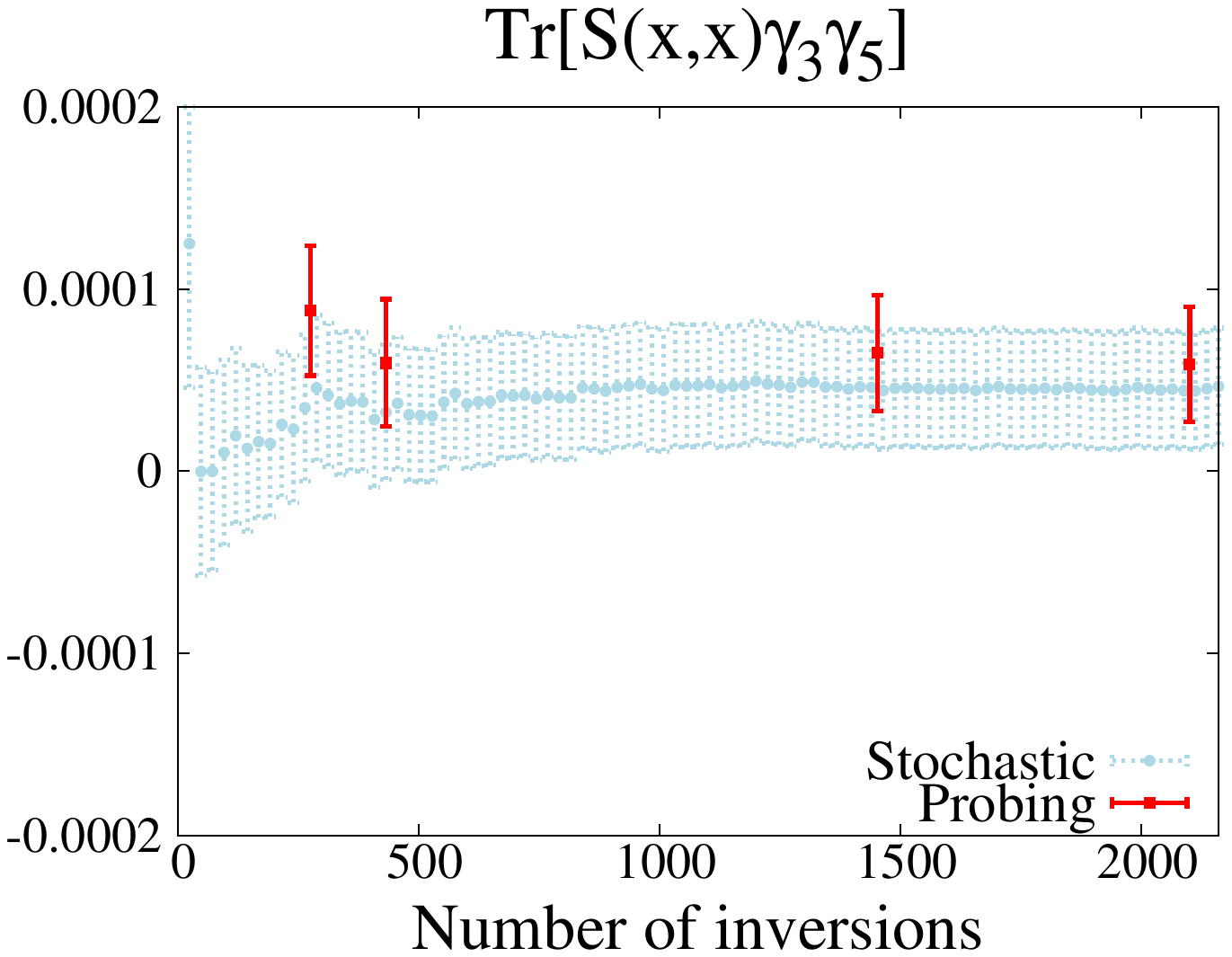}
      \end{center}
    \end{minipage}

\caption{Traces of closed loop propagators for the $16^4$ lattice as a function of computational cost. Red (solid) bars correspond to the probing method for $p=2,\dots,5$. The results of the stochastic technique are denoted by blue (dashed) bars.}
\label{fig:app:collect_IV}
\end{figure*}
In the case of the $16^4$ lattice, the computation of exact propagators is too costly, and we limit ourselves to compare the results from probing and SVS. This is summarized in Figure~\ref{fig:app:collect_IV}. Small deviations from zero in the cases $\Gamma=\gamma_5,\gamma_3\gamma_5$ can probably be ascribed to incomplete thermalization of these runs. In this case it is clear that the error is almost immediately dominated by gauge noise, and there is no appreciable difference between the precision of the two techniques. Thus, increasing the number of inversions on each gauge configuration by either augmenting $p$ or the number of hits $N_{\rm r}^{}$ is not expected to reduce the variance until significantly higher statistics are achieved.
Using the same lattices and setup, we have also explored the behavior of the disconnected contribution to the two-point correlation function in eq.~(\ref{eq:full_2pt_fct}). In this case we again observe that the dominant source of fluctuations is the gauge noise. This suggests that the main priority in a study of e.g. singlet meson masses should be a high-statistics computation, before variance reduction techniques aimed at closed propagators acquire a relevant role.
As an illustration, Fig.~\ref{fig:2p_singlet} shows results for the
disconnected contribution to the two-point function
of the zeroth component of the singlet axial current in our $16^4$ lattice.

\subsection{Eye diagrams for $K\to\pi$}\label{sct:eye}

\begin{figure}[t!]
\centering
\includegraphics[width=6.0cm]{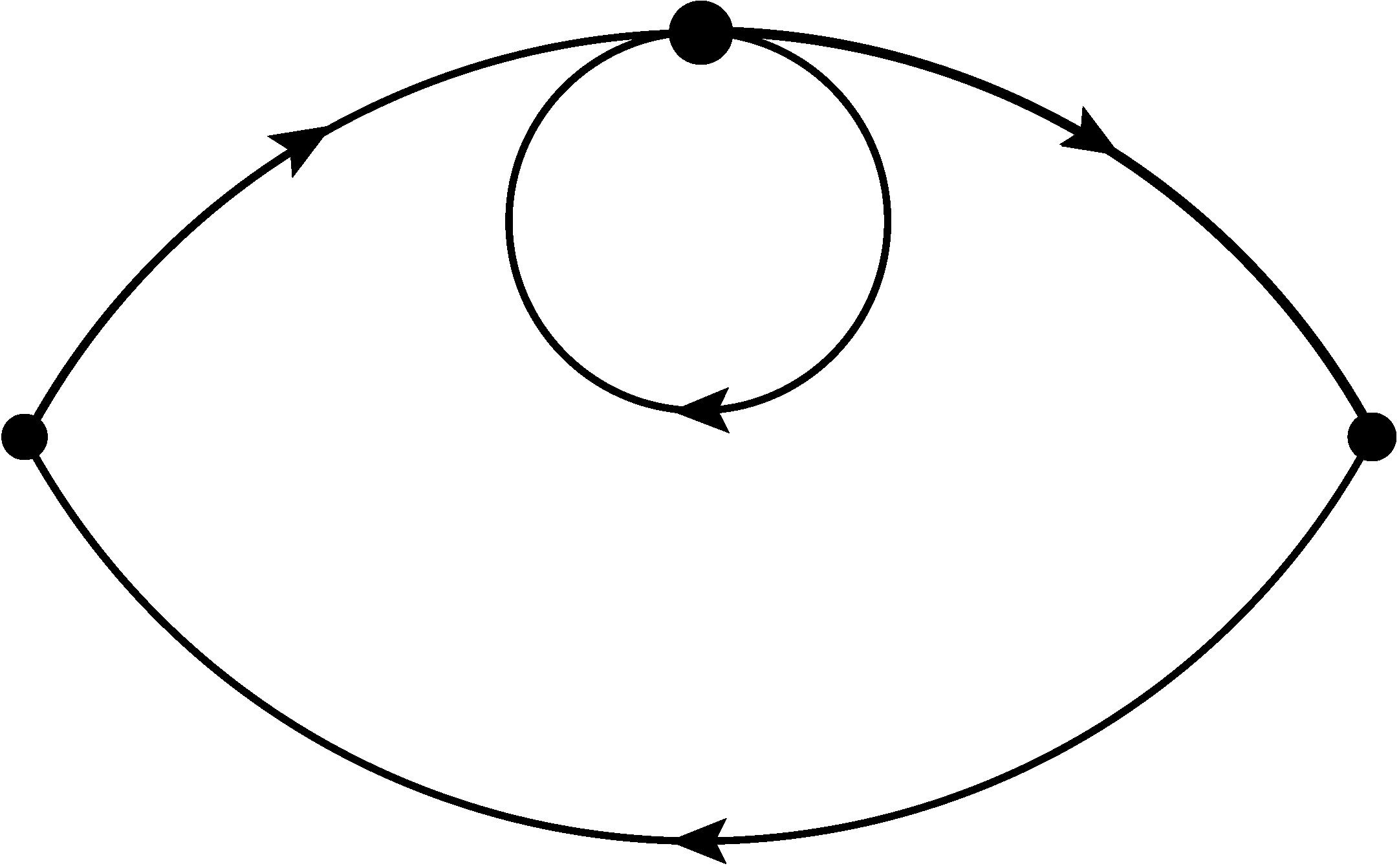}
\caption{Eye diagram occurring in $K\to\pi$ transitions.}\label{fig:eye}
\end{figure}

%
In this section the propagator techniques are applied to an ongoing project to study the role of the charm quark in the explanation of the $\Delta I = 1/2$ rule~\cite{Hartmut,prl,zm}. Our physics results are
presented in a companion paper~\cite{physpaper}; preliminary results from this study have also been reported in Ref.~\cite{proceedings}.
Our strategy is to keep the charm quark as an active degree of freedom in the $\Delta S=1$ effective weak Hamiltonian for $K\to\pi\pi$ decays, and study the charm quark mass $m_c$ dependence of the low-energy constants that control physical amplitudes within a chiral effective description of the decays. The advantage of employing the effective description is that the physics can be addressed by computing $K\to\pi$ transition amplitudes, which avoids some of the difficulties of dealing directly with $K\to\pi\pi$ transitions in Euclidean spacetime. By studying the unphysical GIM limit $m_c=m_u$, where the charm has the same mass as the up quark, intrinsic low-energy effects can be isolated; then by taking the charm mass towards its physical value $m_c \gg m_u$ the impact of having a large mass difference can be understood.
Overlap fermions are used in the lattice computation, since preserving an exact chiral symmetry is
crucial in order to avoid complicated renormalization problems.
A more detailed description of this strategy can be found in Refs.~\cite{Hartmut,prl}.
$K\to\pi$ amplitudes are extracted from the large Euclidean time behavior of three-point correlation functions of $\Delta S=1$ operators with kaon and pion interpolating operators, for which left-handed currents are chosen.\footnote{This avoids technical complications related to exact zero modes of the lattice Dirac operator.}
In the GIM limit, this only leads to Wick contractions which do not involve closed quark propagators (commonly dubbed ``eight diagrams'').
The main technical difficulty when moving to $m_c \neq m_u$ is the appearance of additional contractions in the form of so-called ``eye diagrams'' or ``penguin contractions'', depicted in Figure~\ref{fig:eye}. The corresponding Wick contractions have the form
\begin{align}
C_{\rm eye}^{\rm con}(x^0-z^0,y^0-z^0)=&\left\langle
\text{tr}\left\lbrace  \gamma_{\mu}P_{-}S_{s}(z,x)\gamma_{0}P_{-}S_{u}(x,y)\gamma_{0}P_{-}S_{d}(y,z)\gamma_{\mu}P_{-}S_{u-c}(z,z)\right\rbrace\right\rangle_{\rm G} \label{eqn:eye_con},\\
C_{\rm eye}^{\rm dis}(x^0-z^0,y^0-z^0)=&\left\langle \text{tr}\left\lbrace \gamma_{\mu}P_{-}S_{s}(z,x)
\gamma_{0}P_{-}S_{u}(x,y)\gamma_{0}P_{-}S_{d}(y,z)\right\rbrace\text{tr}\left\lbrace\gamma_{\mu}P_{-}S_{u-c}(z,z)  \right\rbrace\right\rangle_{\rm G}\label{eqn:eye_dis},
\end{align}
where subscripts in quark propagators refer to quark flavors, $P_-=(\mathbf{1}-\gamma_5)/2$, and $S_{u-c}=S_u-S_c$.
From now on we will assume a kinematics where the light up, down, and strange quarks are mass-degenerate.
We will refer to $C_{\rm eye}^{\rm con}$ and $C_{\rm eye}^{\rm dis}$ as ``color-connected'' and ``color-disconnected'', respectively. In order to extract the physical amplitudes, it is convenient to define the quantities
\begin{align} 
E_{}^{-}(x^0-z^0,y^0-z^0) &= \frac{C_{\rm eye}^{\rm dis}(x^0-z^0,y^0-z^0) - C_{\rm eye}^{\rm con}(x^0-z^0,y^0-z^0)}{C(x^0-z^0)C(y^0-z^0)}\label{eqn:ratio_res_I},\\[1.0ex]
E_{}^{+}(x^0-z^0,y^0-z^0) &= \frac{C_{\rm eye}^{\rm dis}(x^0-z^0,y^0-z^0) + C_{\rm eye}^{\rm con}(x^0-z^0,y^0-z^0)}{C(y^0-z^0)C(y^0-z^0)}\label{eqn:ratio_res_II},
\end{align}
where $C$ is a flavor non-singlet two-point function of left-handed currents involving light quarks with mass $m_\ell=m_u=m_d=m_s$, viz.
\begin{equation}
C(x^0-y^0) = -\left\langle \text{tr}\left\{\gamma_{0}^{}P_{-}^{}S_{\ell}(x,y)\gamma_{0}^{}P_{-}^{}S_{\ell}(y,x)\right\}\right\rangle.
\label{eq:2pt_normalization}
\end{equation}
Note that the fact that correlation functions only depend on Euclidean time differences is a consequence of full Poincar\'e invariance. On the lattice this has to be enforced by suitable integrations over space and/or time coordinates at operator insertion points, which is fully possible only with all-to-all propagators. When the latter are not available, estimators involving fixed operator locations have to be employed.
It has been known for a long time that eye diagrams suffer from a severe signal-to-noise problem when the four-fermion operator insertion is set to a fixed spacetime position $z$. Integration over $\vec z$ is expected to result in a variance reduction proportional to the spatial volume, but it requires constructing (the difference of) up and charm closed propagators for all points in space. This adds up to the already high computational effort
required by overlap fermions, resulting in a very demanding setup, the feasibility of which relies
on developing an optimal methodology.
A straightforward possibility is to apply LMA, cf. section~\ref{sct:lma}. This will result in a splitting of the three-point functions into a total of sixteen contributions, since each of the four quark lines in Figure~\ref{fig:eye} are split into $S^{\rm l}+S^{\rm h}$. They can be collectively labeled as
\begin{equation}
\label{eqn:lma_3p}
C_{\rm eye} = C^{\rm llll} + C^{\rm hlll} + C^{\rm hhll} + C^{\rm hhhl} + C^{\rm hhhh}\,,
\end{equation}
with $C^{\rm hlll}$ and $C^{\rm hhhl}$ containing four different contributions and $C^{\rm hhll}$ containing six different contributions. Whenever only low-mode propagators meet at $z$, it will be possible to integrate over space at $z$, thus averaging a significant fraction of the contributions involving large fluctuations. This was indeed shown in Ref.~\cite{prl,procsdublin} to have a large variance reduction effect in the case of eight diagrams, where however the signal-to-noise problem is less severe. Indeed, preliminary studies revealed that LMA alone, in the form employed in Ref.~\cite{prl}, is not enough to obtain a signal for eye diagrams~\cite{unpublished}. Here we will study the hybrid strategy of combining LMA and SVS and/or probing (in closed propagator contributions) for the estimation of the high-mode part of quark propagators, and devise optimal methods for variance reduction.

%
%
\subsubsection{LMA + SVS}\label{sbsbsct:SVS_EYE_DIAGRAMS_DILUTION}
%

%
In this section we will combine LMA with diluted SVS in order to compute the high-mode propagators $S^{\rm h}$. In the case of eye diagrams two different types of propagators are involved. On the one hand, there are the loop propagators which start and end at the same lattice site, such that the space-diagonal elements of the propagator matrix are required;
on the other hand, there are propagators connecting different spacetime locations outside the loop. In the following the latter will be referred to as leg propagators.
Since from the computational point of view the structure of loop and leg propagators is very different, different dilution schemes can in principle be chosen for each of them, in order to optimize the total signal.
Computations are carried out with overlap fermions~\cite{OVERLAPP_1,OVERLAPP_2}, generated with the Wilson plaquette action at $\beta= 5.8485$.  The latter corresponds to a lattice spacing given in terms of the Sommer parameter $r_{0}^{}\approx 0.5$~fm by $a/r_{0}^{}\simeq 0.237$~\cite{Necco:2001xg}.
The Neuberger-Dirac operator is constructed and inverted using the techniques described in Ref.~\cite{oveps}. In order to compare the computational cost of dilution schemes, they are tuned by fixing the least common multiple number of inversions, i.e. the number of inversions is identical for all stochastic results. This requires balancing dilution and hit number --- the lesser diluted, the higher the number of hits. Time dilution is applied by default in all cases. We will also compare the results from SVS with those from LMA making optimal use of extended propagators, cf. section~\ref{sct:lma}.

\paragraph{Dilution schemes}

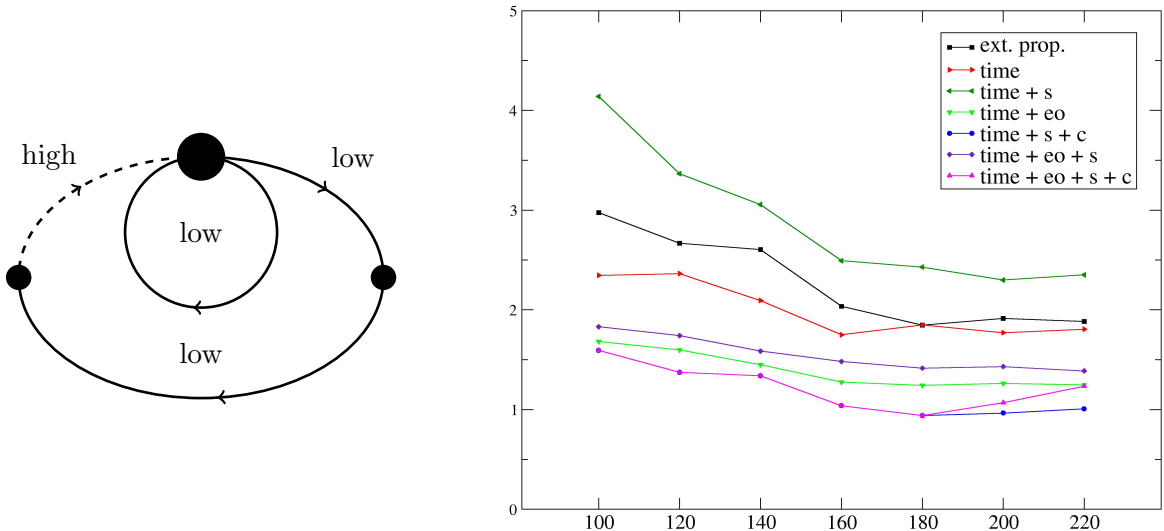
\begin{figure}[t!]
\vspace*{-20mm}
	\hspace*{-0mm}
    \begin{minipage}[c]{.5\linewidth}
    \vspace{40mm}
\begin{tikzpicture}[line width=1. pt, scale=1.0]
 \begin{scope}
		\vphantom{\draw[
        decoration={markings, mark=at position 0.75 with {\arrow{<}}},
        style={dashed},postaction={decorate}
        ] (0,-1) circle (1.0);}
		\draw[
        decoration={markings, mark=at position 0.75 with {\arrow{<}}},
        postaction={decorate}
        ] (0,4) circle (1.0);
			\node at (0,4) {low};
			\node at (0,1) {};
			\node at (2,5) {low};
			\node at (-2,5) {high};
			\node at (0,2.4) {low};
			\draw[scalarbar]  (0, 5) arc (90:180:2.4 and 1.6);	
			\draw[fermionbar]  (-2.4, 3.4) arc (180:360:2.4 and 1.6);	
			\draw[fermion]     (0, 5) arc (270:360:2.4 and -1.6);	
		  \draw[fill=black] (-2.4,3.4) circle (.15cm);
		  \draw[fill=black] (2.4,3.4)  circle (.15cm);			
			\draw[fill=black] (0,5) circle (.3cm);
				\end{scope}	
\end{tikzpicture}
    \end{minipage}
    	\hspace*{-20mm}
    \begin{minipage}[c]{.5\linewidth}
      \begin{center}
      \includegraphics[width=95mm]{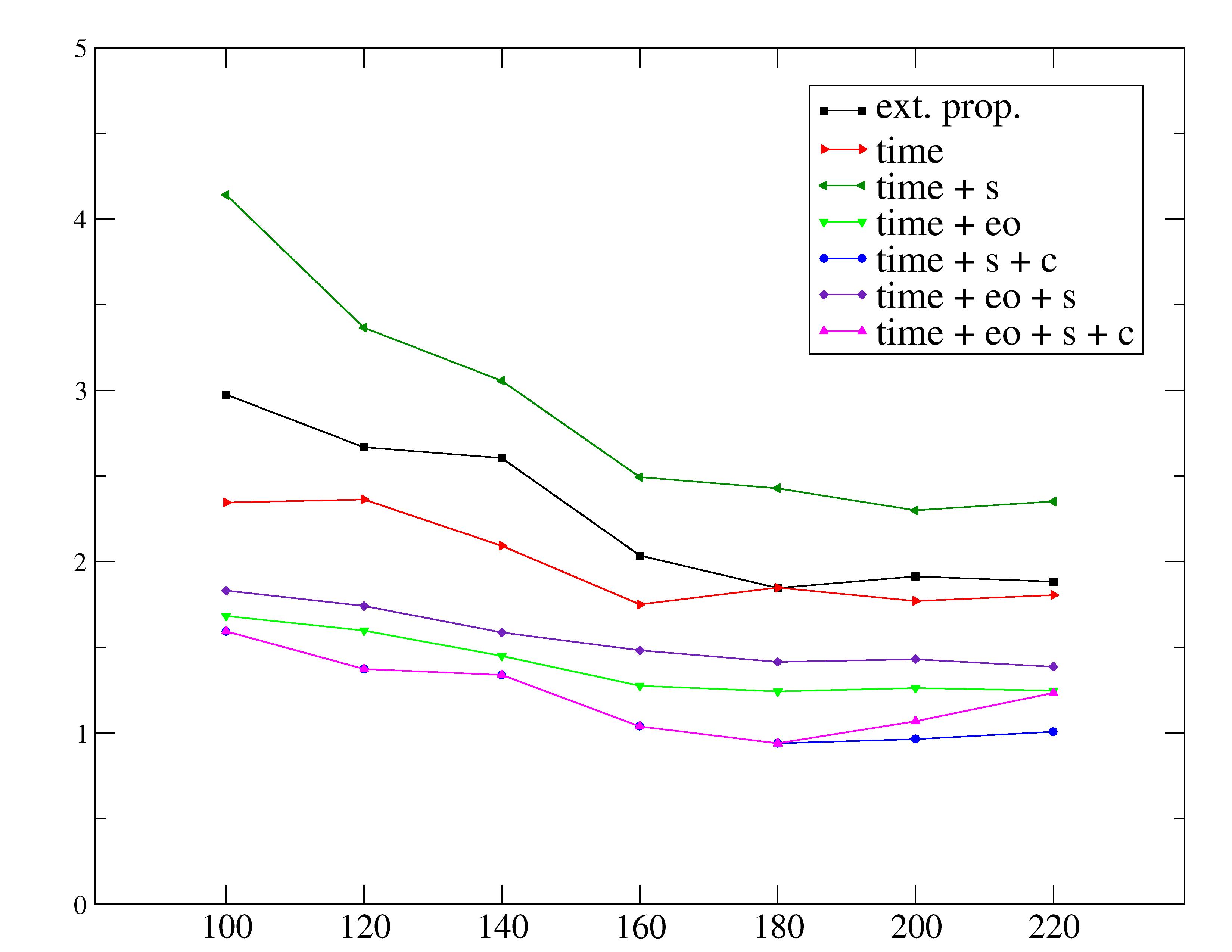}
      \end{center}
    \end{minipage}

\vspace*{-20mm}
\caption{Comparison of dilution schemes for the high-mode leg propagator. Dilution schemes are combinations of time, spin (s), color (c) and even-odd (eo) dilution. As a benchmark the result of computing the leg propagator via an extended propagator is shown. Shown is the value of this specific contribution to the diagram (in lattice units) as a function of the number
of configurations taken for the gauge average.}
\label{fig:DIL_OUTER_LOOP}
\end{figure}

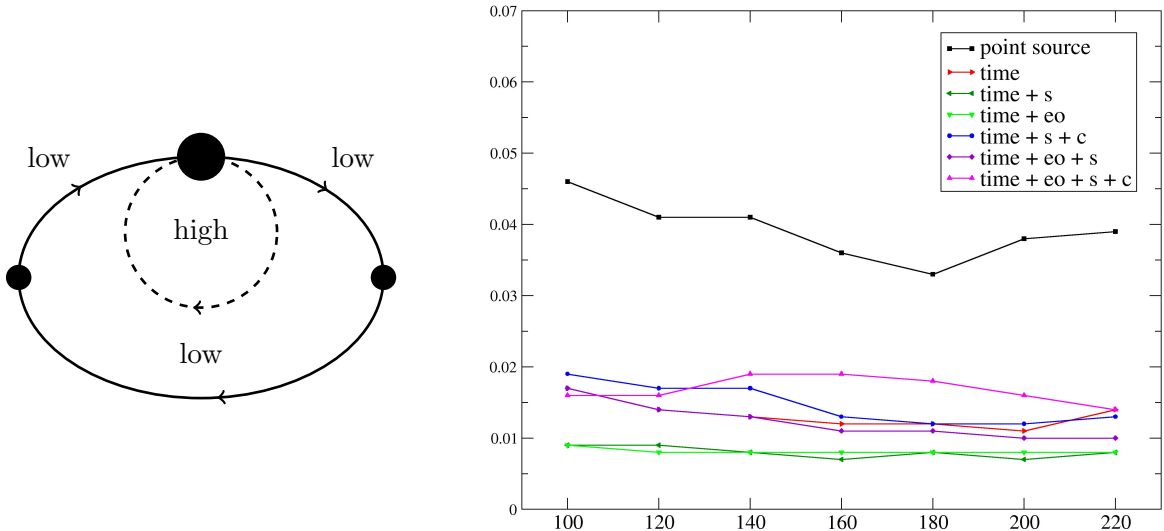
\begin{figure}[t!]
\vspace*{-20mm}
	\hspace*{-0mm}
    \begin{minipage}[c]{.5\linewidth}
    \vspace{40mm}
\begin{tikzpicture}[line width=1. pt, scale=1.0]
 \begin{scope}
		\vphantom{\draw[
        decoration={markings, mark=at position 0.75 with {\arrow{<}}},
        style={dashed},postaction={decorate}
        ] (0,-1) circle (1.0);}
		\draw[
        decoration={markings, mark=at position 0.75 with {\arrow{<}}},
        style={dashed},postaction={decorate}
        ] (0,4) circle (1.0);
			\node at (0,4) {high};
			\node at (0,1) {};
			\node at (2,5) {low};
			\node at (-2,5) {low};
			\node at (0,2.4) {low};
			\draw[fermionbar]  (0, 5) arc (90:180:2.4 and 1.6);	
			\draw[fermionbar]  (-2.4, 3.4) arc (180:360:2.4 and 1.6);	
			\draw[fermion]     (0, 5) arc (270:360:2.4 and -1.6);	
		  \draw[fill=black] (-2.4,3.4) circle (.15cm);
		  \draw[fill=black] (2.4,3.4)  circle (.15cm);			
			\draw[fill=black] (0,5) circle (.3cm);
				\end{scope}	
\end{tikzpicture}
    \end{minipage}
    	\hspace*{-20mm}
    \begin{minipage}[c]{.5\linewidth}
      \begin{center}
      \includegraphics[width=95mm]{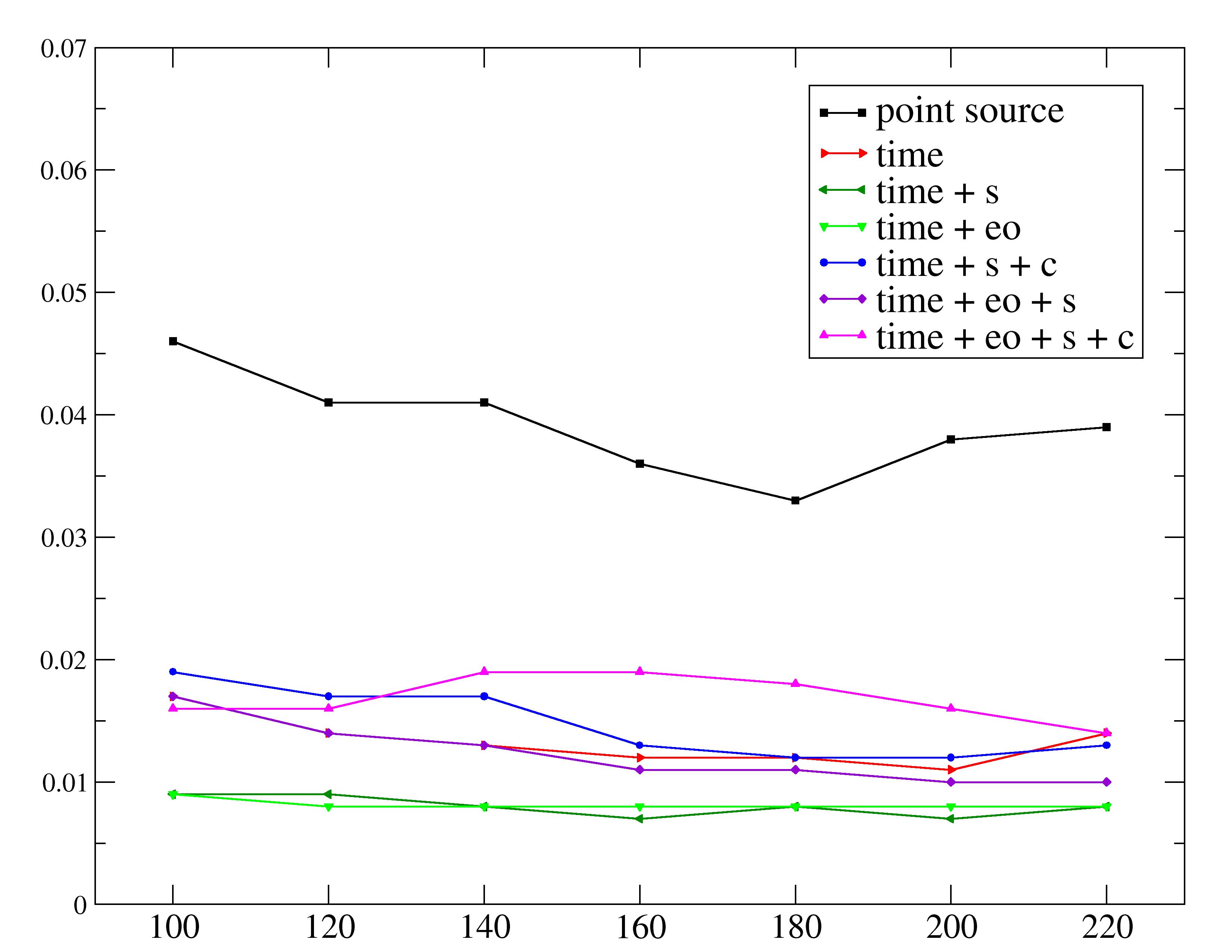}
      \end{center}
    \end{minipage}

\vspace*{-20mm}
\caption{Comparison of dilution schemes for the high-mode  loop propagator. Dilution schemes are combinations of time, spin (s), color (c) and even-odd (eo) dilution. For comparison the result of computing the loop propagator via a standard point source is shown (extended propagators are not applicable here). Shown is the value of this specific contribution to the diagram (in lattice units) as a function of the number
of configurations taken for the gauge average.}
\label{fig:DIL_INNER_LOOP}
\end{figure}

We will first discuss results obtained with 220 quenched configurations on a volume
$Va_{}^{-4}=16_{}^{4}$, with light quark mass $am_{\ell}^{}=0.02$ and charm mass $am_{c}^{}=0.04$. We will focus on the two particular contributions to eye diagrams shown in the left panels of Figures~\ref{fig:DIL_OUTER_LOOP},\ref{fig:DIL_INNER_LOOP}. These are suitable benchmark cases, since three of the four involved propagators are constructed out of low-modes and are thus exact all-to-all propagators. Only a single high-mode propagator has to be estimated stochastically, and for this reason the introduced intrinsic stochastic noise in different dilution schemes can be compared directly.
The right panels of Figures~\ref{fig:DIL_OUTER_LOOP},\ref{fig:DIL_INNER_LOOP} show the absolute error (of the value at $t=x_{0}^{}-z_{0}^{}=10a$ for suitable chosen and fixed $y_{0}^{}-z_{0}^{}$) times the square root of the number of configurations in the gauge average.
%

%

%
For the leg propagator dilution in spin$+$color turns out to be favorable, followed by spin$+$ color$+$even-odd and even-odd only. Except for dilution in spin only, the use of stochastic volume sources is at least as good as the use of extended propagators. This is remarkable, since the finding implies that a single stochastic all-to-all propagator in the legs reaches the same accuracy as an ``exact'' extended propagator. 
The computational effort of the stochastic technique is significantly higher, though. 
Regarding the loop propagator, dilution in spin or even-odd dilution seem to be advantageous. Most importantly, thanks to being able to average over the coordinate of the loop origin, the noise can be reduced substantially by combining LMA with SVS for arbitrary dilution schemes.
A crucial finding is that identical random source vectors have to be used for the stochastic estimation of both the up and charm quark loop, in order to avoid artificial stochastic noise in the $u-c$ difference, cf. eqs.~(\ref{eqn:eye_con},\ref{eqn:eye_dis}); otherwise no signal is observed.
Moreover, it can be seen that in both cases the expected rate of convergence with increasing number of configurations taken in the gauge average is observed --- i.e. for a large number of gauge configurations $N_{\text{cfg}}^{}$ the error decreases with $\sqrt{N_{\text{cfg}}^{}}$. This shows that the Monte Carlo integration is well-behaved and there is no evidence of large fluctuations that may destroy the average.
Finally, we remark that, while it is possible to choose different dilution schemes for leg and loop propagators, in practice it is convenient to select the same scheme. In this way, the loop propagator of the up quark can be reused as an independent stochastic hit for the leg propagator whenever it is not required for the loop. In terms of computational cost for a given level of signal, this is advantageous even
if the dilution scheme is not optimal in one of the two cases.
%

%

\paragraph{Stochastic noise reduction}

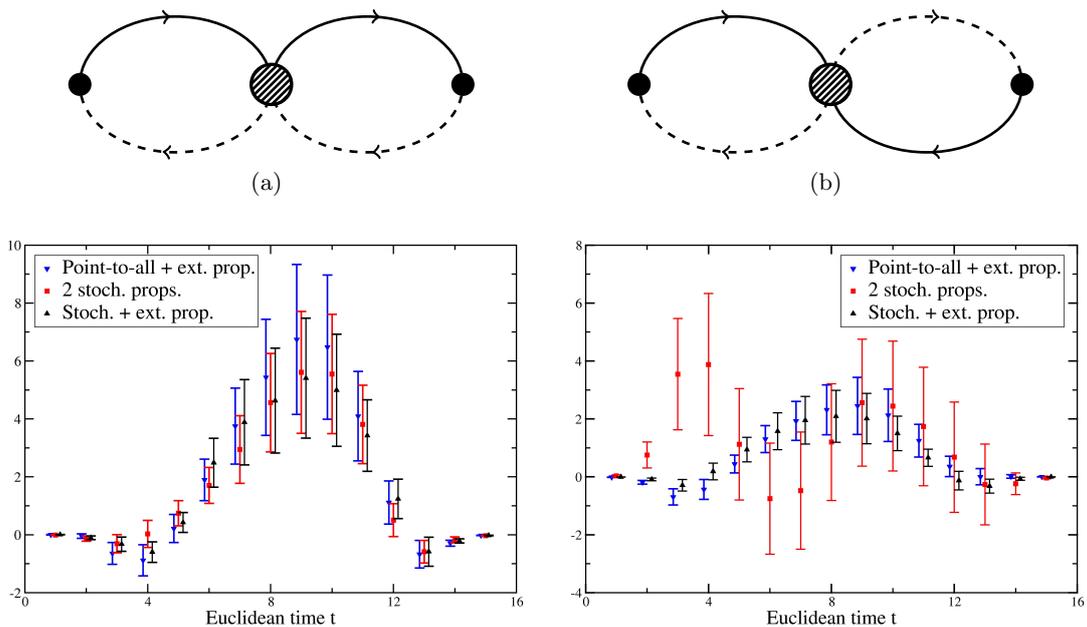
\begin{figure}[t!]
\centering
\subfigure[]{
\begin{tikzpicture}[line width=1. pt, scale=0.9]
   \begin{scope}
			\draw[fermionbar]  (0, 0) arc (0:180:1.4 and 1);	
			\draw[scalarbar] (0, 0) arc (180:360:1.4 and 1);	
			\draw[scalar]     (0, 0) arc (0:180:1.4 and   -1);	
			\draw[fermion]     (0, 0) arc (180:360:1.4 and -1);	
		  \draw[fill=black] (-2.8,0) circle (.15cm);
		  \draw[fill=black] (2.8,0)  circle (.15cm);			
			\draw[fill=black] (0,0) circle (.3cm);
		  \draw[fill=white] (0,0) circle (.29cm);
		  \begin{scope}
	    	\clip (0,0) circle (.3cm);
	    	\foreach \x in {-.9,-.8,...,.3}
				\draw[line width=1 pt] (\x,-.3) -- (\x+.6,.3);
	  	\end{scope}
   \end{scope}
\end{tikzpicture}\label{sb:fig:hhll}
}
\hspace{1.5cm}
\subfigure[]{
\begin{tikzpicture}[line width=1. pt, scale=0.9]
   \begin{scope}
			\draw[fermionbar]      (0, 0) arc (0:180:1.4 and 1);	
			\draw[fermionbar]     (0, 0) arc (180:360:1.4 and 1);	
			\draw[scalar]   (0, 0) arc (0:180:1.4 and   -1);	
			\draw[scalar]     (0, 0) arc (180:360:1.4 and -1);	
		  \draw[fill=black] (-2.8,0) circle (.15cm);
		  \draw[fill=black] (2.8,0)  circle (.15cm);			
			\draw[fill=black] (0,0) circle (.3cm);
		  \draw[fill=white] (0,0) circle (.29cm);
		  \begin{scope}
	    	\clip (0,0) circle (.3cm);
	    	\foreach \x in {-.9,-.8,...,.3}
				\draw[line width=1 pt] (\x,-.3) -- (\x+.6,.3);
	  	\end{scope}
   \end{scope}
\end{tikzpicture}\label{sb:fig:hlhl}
}
\vspace{0.25cm}
\subfigure{
\includegraphics[width=7.0cm]{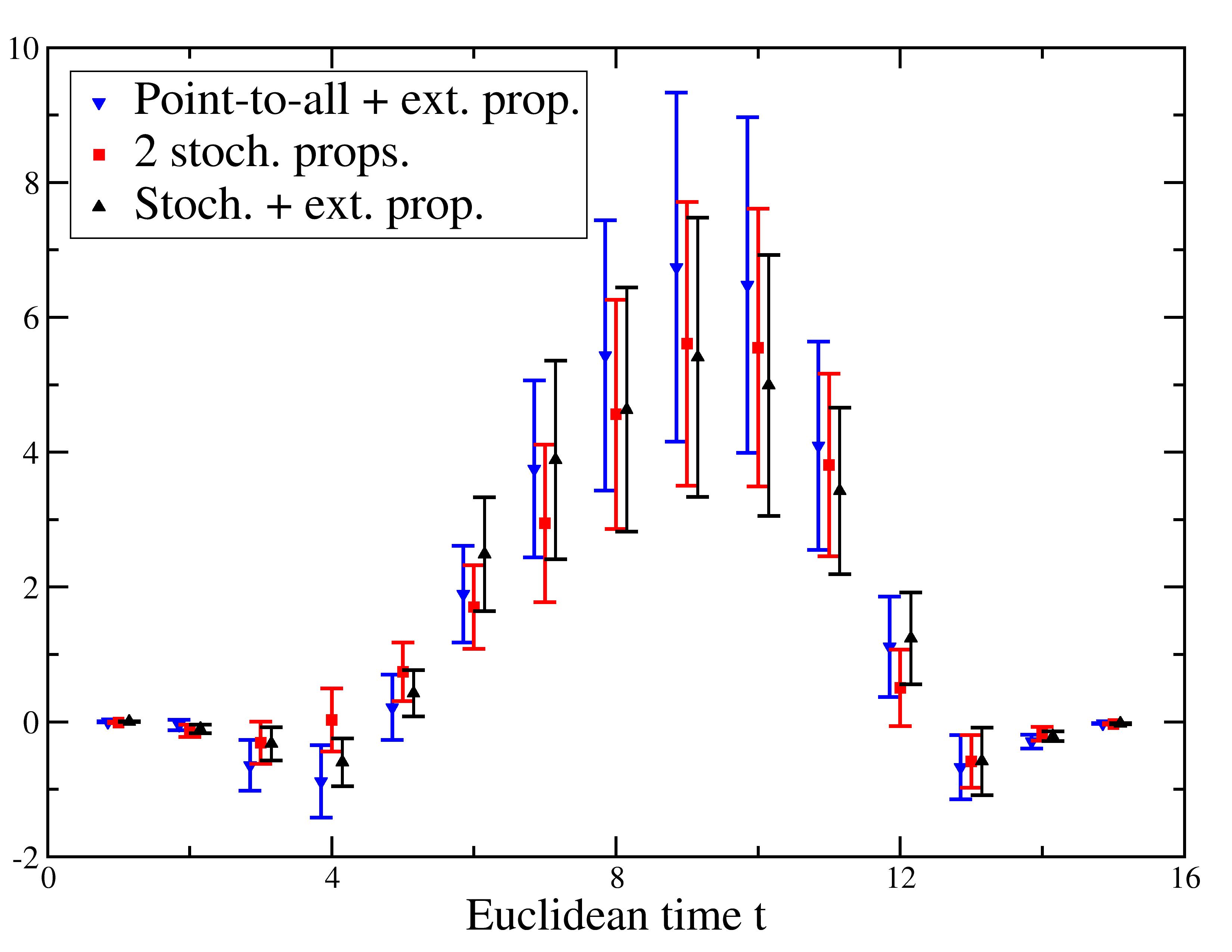}\label{fig:eight_good}
}
\subfigure{
\includegraphics[width=7.0cm]{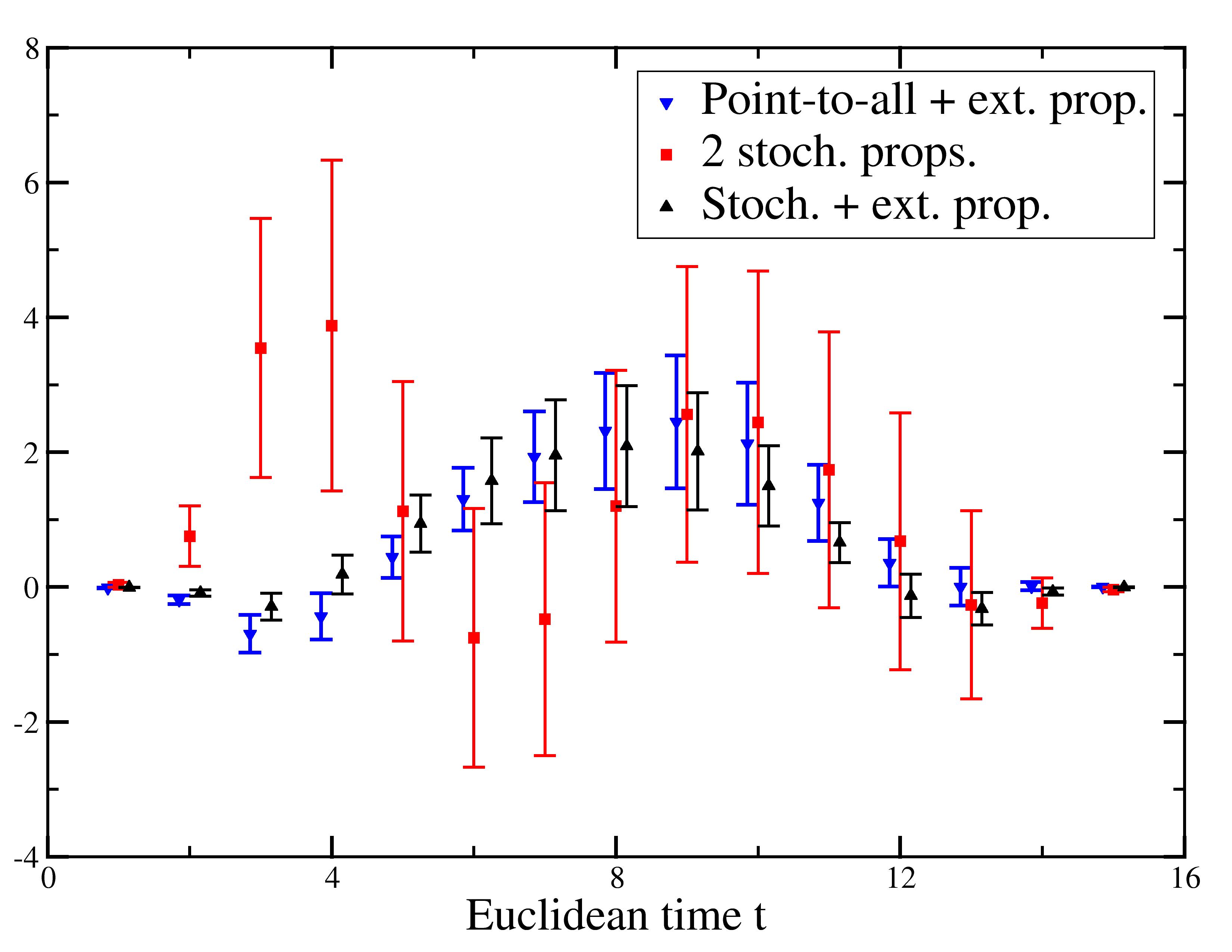}\label{fig:eight_bad}
}
\caption{SVS applied to eight diagrams. Top: Considered contributions (low-mode propagators indicated by solid lines, high-mode propagators are dashed); the different relative location of high- and low-mode contributions means that the respective propagators appear in different orderings within spin traces, cf. e.g. Appendix~B of~\cite{physpaper}. Bottom: Euclidean time dependence of two color-connected contributions to the correlators. Different methods for the computation of the two high-mode propagators are compared: point-to-all + extended propagators (blue down-triangles),  stochastic estimates for $S^{\rm h}$ (red squares), and combination of stochastic and extended propagators (black up-triangles). Shown is the value of this specific contribution to the diagram (in lattice units) as a function of the number
of configurations taken for the gauge average. The points are slightly displaced for better readability.}
\label{fig:EIGHT_GOOD_VS_BAD}
\end{figure}

The level of stochastic noise introduced by SVS can depend on the details of how
random source and solution vectors are arranged in the computation of a specific
contraction.
In order to understand this issue in our context,
we will study it in the well-controlled case of eight diagrams,
using relatively cheap small-volume runs.
The latter are carried out on a $16\times 8_{}^{3}$ lattice, again at $\beta=5.8485$.
Our ensemble consists of 205 quenched configurations. Four stochastic source vectors are diluted in time and spin, while LMA is implemented with $N_{\rm low}=20$ low-modes.
The Wick contractions leading to eight diagrams are of the form
\begin{align}
C_{\rm eight}^{\rm con}(x^0-z^0,y^0-z^0)=&\left\langle
\text{tr}\left\lbrace  \gamma_{\mu}P_{-}S_{s}(z,x)\gamma_{0}P_{-}S_{u}(x,z)\gamma_{\mu}P_{-}S_{u}(z,y)\gamma_{0}P_{-}S_{d}(y,z)\right\rbrace\right\rangle_{\rm G} \label{eqn:eight_con},\\
C_{\rm eight}^{\rm dis}(x^0-z^0,y^0-z^0)=&\left\langle \text{tr}\left\lbrace \gamma_{\mu}P_{-}S_{s}(z,x)
\gamma_{0}P_{-}S_{u}(x,z)\right\rbrace\text{tr}\left\lbrace\gamma_{\mu}P_{-}S_{u}(z,y)\gamma_{0}P_{-}S_{d}(y,z)  \right\rbrace\right\rangle_{\rm G}\label{eqn:eight_dis}.
\end{align}
We then combine them and normalize the results with a suitable product of two-point functions in the
same way as in Eqs.~(\ref{eqn:ratio_res_I},\ref{eqn:ratio_res_II})
We will be specifically interested in exploring contributions with two high-mode propagators,
illustrated by the diagrams in the upper panel of Figure~\ref{sb:fig:hhll} and \ref{sb:fig:hlhl}.
We consider two computational procedures: in the first one, both occurrences of $S^{\rm h}$ are estimated
stochastically; in the second, the stochastic estimation is combined with extended propagators.
In Figure~\ref{fig:EIGHT_GOOD_VS_BAD} both are compared with the result of treating contributions
from $S^{\rm h}$ via extended propagators.
For the correlator on the l.h.s., the use of two independent stochastic estimates yields the same level of accuracy as combining one stochastic or point-to-all propagator with an additional extended propagator. This is a promising finding, since the latter is an exact propagator which ensures minimal variance. Thus, using stochastic all-to-all propagators is competitive with exact techniques based on extended propagators.
On the contrary, the contribution depicted in Figure~\ref{sb:fig:hlhl} shows a pathology that occurs whenever two stochastic source vectors meet at the same lattice location (here the location of the four-fermion operator insertion). 
The use of an extended propagator instead of a second stochastic estimate allows to solve the noise problem by removing one of the random vectors. 
Alternatively, $\gamma_{5}^{}$-hermiticity may be applied. This allows to interchange the position of solution and source vector. Consequently, only one random vector remains at the position of the four-fermion operator insertion. \footnote{Recently, the RBC/UKQCD collaboration reported on a similar behavior in the context of computing direct $K\to\pi\pi$ decays~\cite{noise_rbc}. In their case, it is observed that having several random vectors at the four-quark operator insertion leads to a bad signal-to-noise ratio.
The suggested solution consists of switching the source and sink coordinates of the
propagators involved using $\gamma_{5}^{}$-hermiticity.}

\paragraph{Exploratory physics runs}

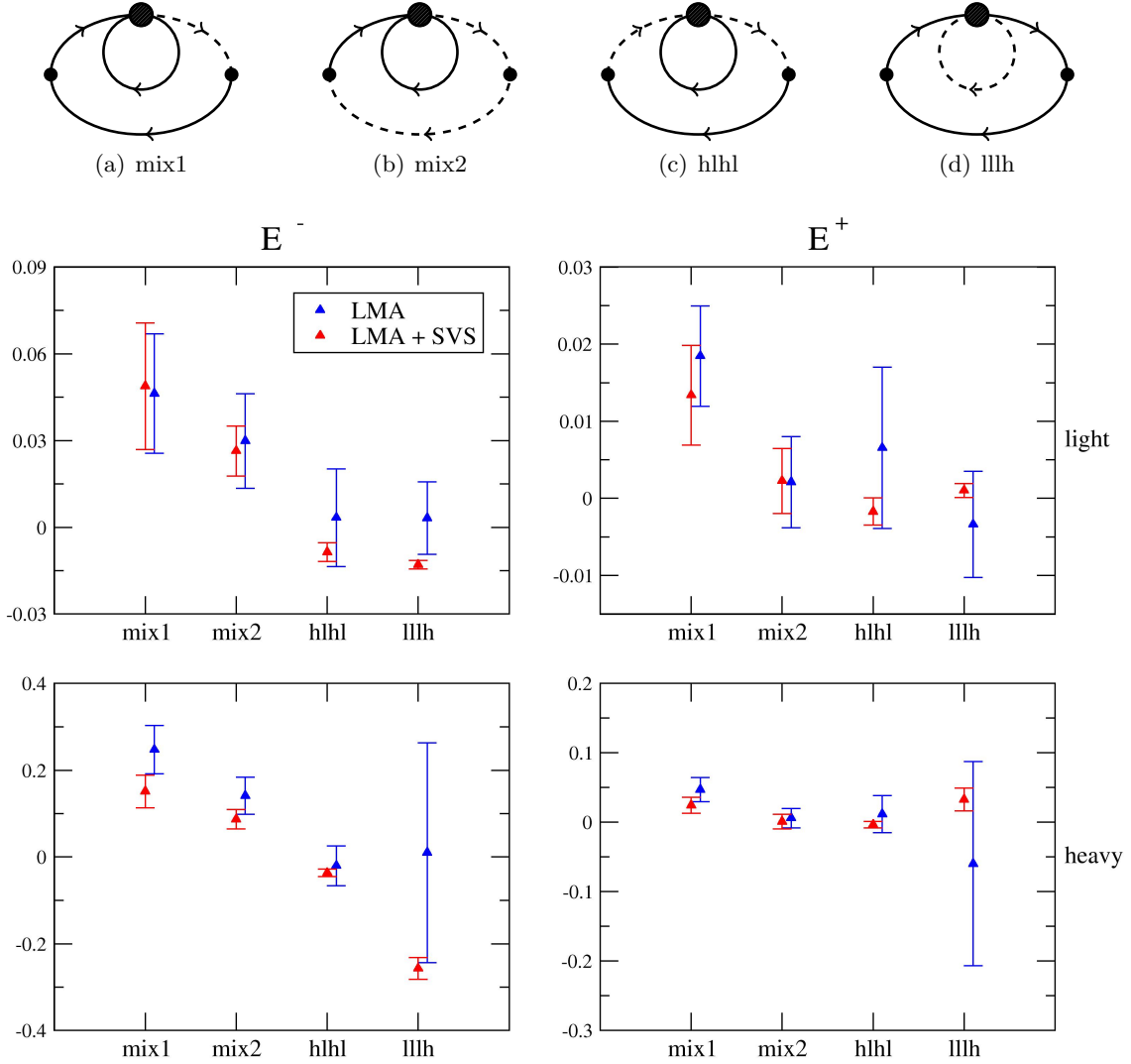
\begin{figure*}[t!]
\begin{center}
\subfigure[mix1]{
\begin{tikzpicture}[line width=1. pt, scale=0.5]
    \begin{scope}
		\draw[
        decoration={markings, mark=at position 0.75 with {\arrow{<}}},
        postaction={decorate}
        ] (0,-1) circle (1.0);
			\draw[fermionbar]  (0, 0) arc (90:180:2.4 and 1.6);	
			\draw[fermionbar]  (-2.4, -1.6) arc (180:360:2.4 and 1.6);	
			\draw[scalar]     (0, 0) arc (270:360:2.4 and -1.6);	
		  \draw[fill=black] (-2.4,-1.6) circle (.15cm);
		  \draw[fill=black] (2.4,-1.6)  circle (.15cm);			
			\draw[fill=black] (0,0) circle (.3cm);
		  \draw[fill=white] (0,0) circle (.29cm);	
			\begin{scope}
	    	\clip (0,0) circle (.3cm);
	    	\foreach \x in {-.9,-.8,...,.3}
				\draw[line width=1 pt] (\x,-.3) -- (\x+.6,.3);
	  	\end{scope}
		\end{scope}	
\end{tikzpicture}\label{fig:mix1}
}
\hspace{0.6cm}
\subfigure[mix2]{
\begin{tikzpicture}[line width=1. pt, scale=0.5]
 \begin{scope}
		\draw[
        decoration={markings, mark=at position 0.75 with {\arrow{<}}},
        postaction={decorate}
        ] (0,-1) circle (1.0);
			\draw[fermionbar]  (0, 0) arc (90:180:2.4 and 1.6);	
			\draw[scalarbar]  (-2.4, -1.6) arc (180:360:2.4 and 1.6);	
			\draw[scalar]     (0, 0) arc (270:360:2.4 and -1.6);	
		  \draw[fill=black] (-2.4,-1.6) circle (.15cm);
		  \draw[fill=black] (2.4,-1.6)  circle (.15cm);			
			\draw[fill=black] (0,0) circle (.3cm);
		  \draw[fill=white] (0,0) circle (.29cm);	
			\begin{scope}
	    	\clip (0,0) circle (.3cm);
	    	\foreach \x in {-.9,-.8,...,.3}
				\draw[line width=1 pt] (\x,-.3) -- (\x+.6,.3);
	  	\end{scope}
		\end{scope}	
\end{tikzpicture}\label{fig:mix2}
}
\hspace{0.6cm}
\subfigure[hlhl]{
\begin{tikzpicture}[line width=1. pt, scale=0.5]
 \begin{scope}
		\draw[
        decoration={markings, mark=at position 0.75 with {\arrow{<}}},
        postaction={decorate}
        ] (0,-1) circle (1.0);
			\draw[scalarbar]  (0, 0) arc (90:180:2.4 and 1.6);	
			\draw[fermionbar]  (-2.4, -1.6) arc (180:360:2.4 and 1.6);	
			\draw[scalar]     (0, 0) arc (270:360:2.4 and -1.6);	
		  \draw[fill=black] (-2.4,-1.6) circle (.15cm);
		  \draw[fill=black] (2.4,-1.6)  circle (.15cm);			
			\draw[fill=black] (0,0) circle (.3cm);
		  \draw[fill=white] (0,0) circle (.29cm);	
			\begin{scope}
	    	\clip (0,0) circle (.3cm);
	    	\foreach \x in {-.9,-.8,...,.3}
				\draw[line width=1 pt] (\x,-.3) -- (\x+.6,.3);
	  	\end{scope}
		\end{scope}	
\end{tikzpicture}\label{fig:mix3_a}
}
\hspace{0.6cm}
\subfigure[lllh]{
\begin{tikzpicture}[line width=1. pt, scale=0.5]
 \begin{scope}
		\draw[
        decoration={markings, mark=at position 0.75 with {\arrow{<}}},
        style={dashed},postaction={decorate}
        ] (0,-1) circle (1.0);
			\draw[fermionbar]  (0, 0) arc (90:180:2.4 and 1.6);	
			\draw[fermionbar]  (-2.4, -1.6) arc (180:360:2.4 and 1.6);	
			\draw[fermion]     (0, 0) arc (270:360:2.4 and -1.6);	
		  \draw[fill=black] (-2.4,-1.6) circle (.15cm);
		  \draw[fill=black] (2.4,-1.6)  circle (.15cm);			
			\draw[fill=black] (0,0) circle (.3cm);
		  \draw[fill=white] (0,0) circle (.29cm);	
			\begin{scope}
	    	\clip (0,0) circle (.3cm);
	    	\foreach \x in {-.9,-.8,...,.3}
				\draw[line width=1 pt] (\x,-.3) -- (\x+.6,.3);
	  	\end{scope}
		\end{scope} 	 
\end{tikzpicture}\label{fig:mix3_b}
}
\\
\subfigure{
\includegraphics[width=15.0cm]{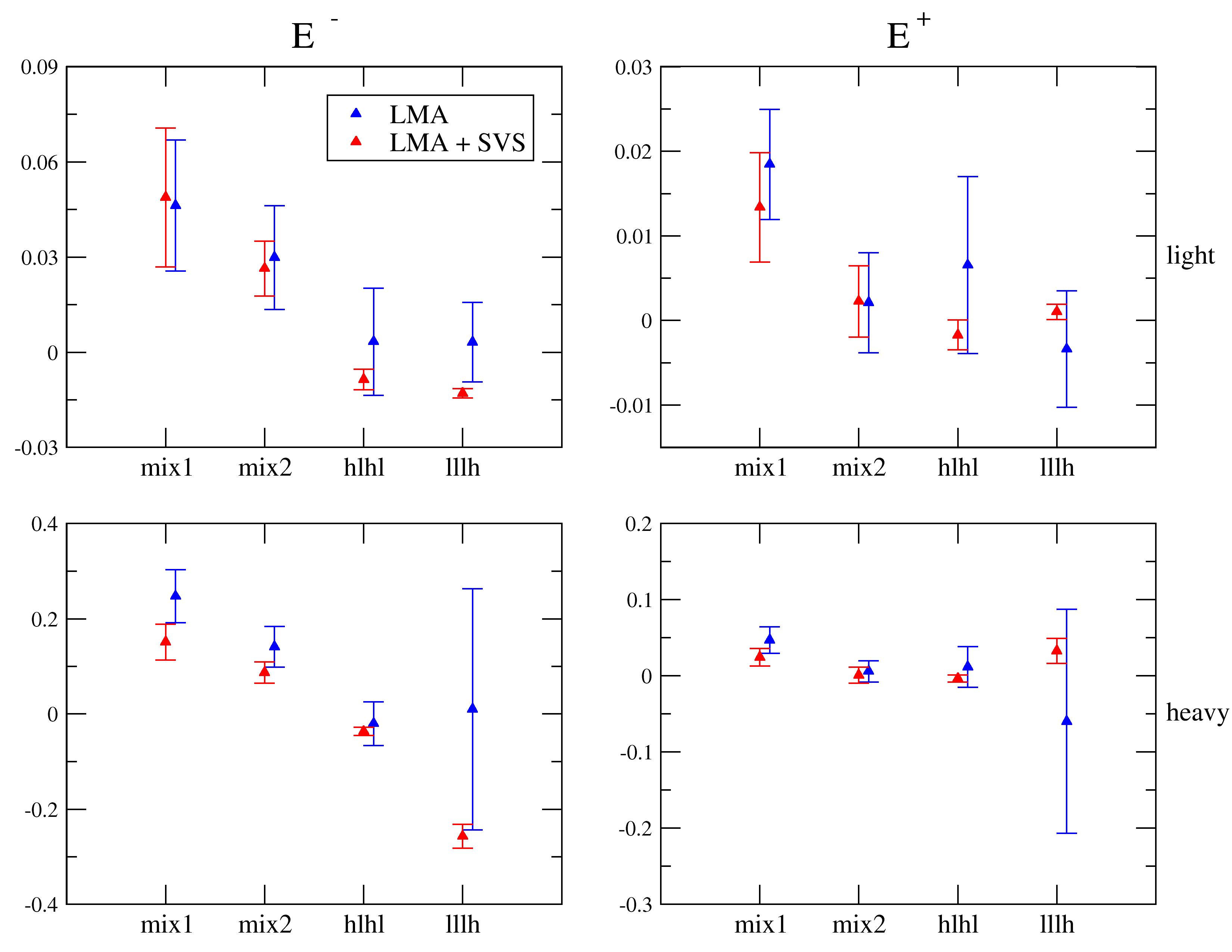}\label{fig:mainz_report_plot}
}
\end{center}
\caption{Top: Selection of contributions to eye diagrams within LMA. Low-mode propagators $S_{}^{\rm l}$ are denoted by solid lines; high-mode propagators $S_{}^{\rm h}$ are dashed. Bottom: Fit results for the considered contributions. The upper (lower) row includes the results of a light (heavy) charm quark while the columns distinguish the ratios $E_{}^{\pm}$.
 Shown is the value of this specific contribution to the diagram (in lattice units) as a function of the number
of configurations taken for the gauge average; note the different scales in each plot.}
\label{fig:SELECTION_EYE_DIAGRAMS}%
\end{figure*}

\begin{figure}[t!]

	\hspace*{-2mm}
    \begin{minipage}{.5\linewidth}
      \begin{center}
      \includegraphics[width=7.5cm]{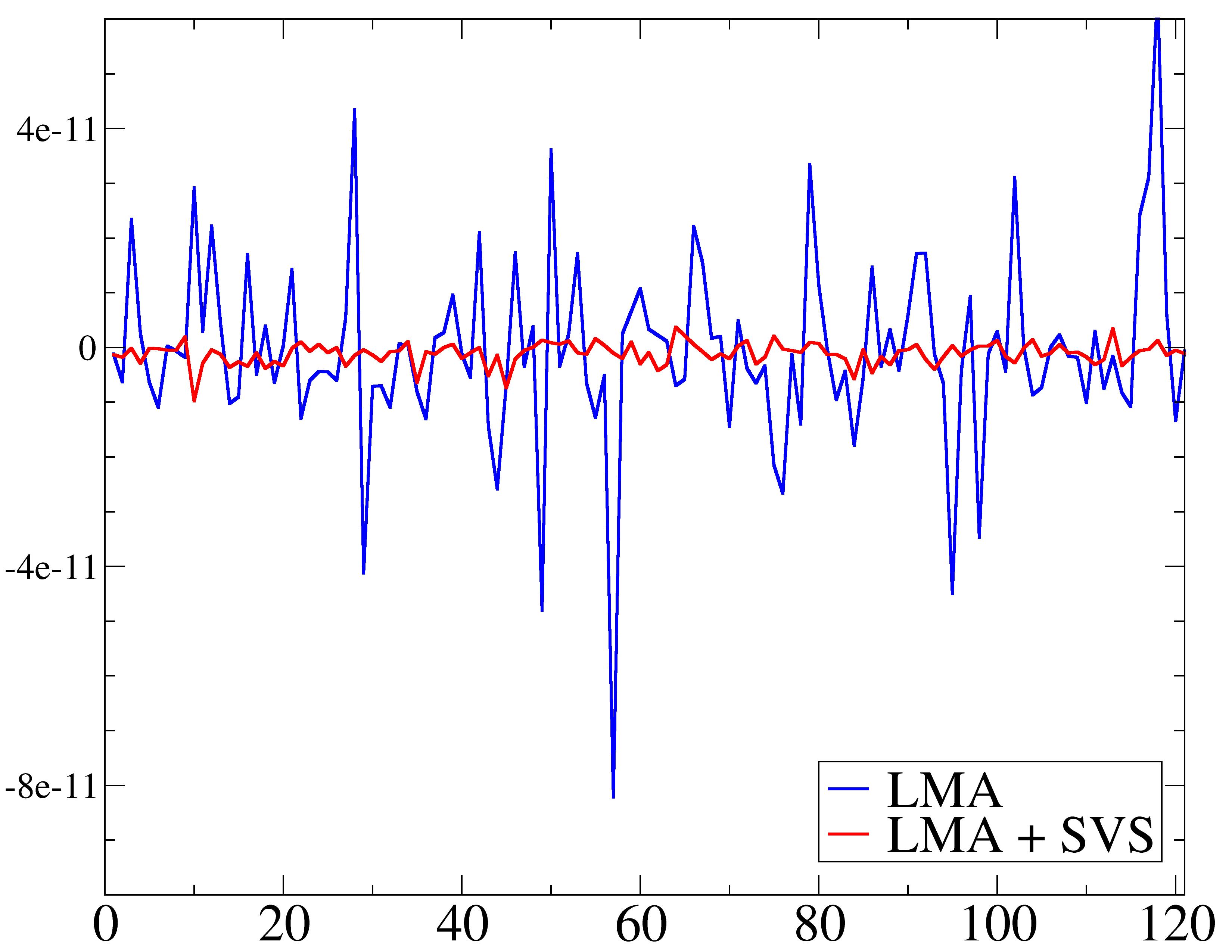}
      \end{center}
    \end{minipage}
    \begin{minipage}{.5\linewidth}
      \begin{center}
      \includegraphics[width=7.5cm]{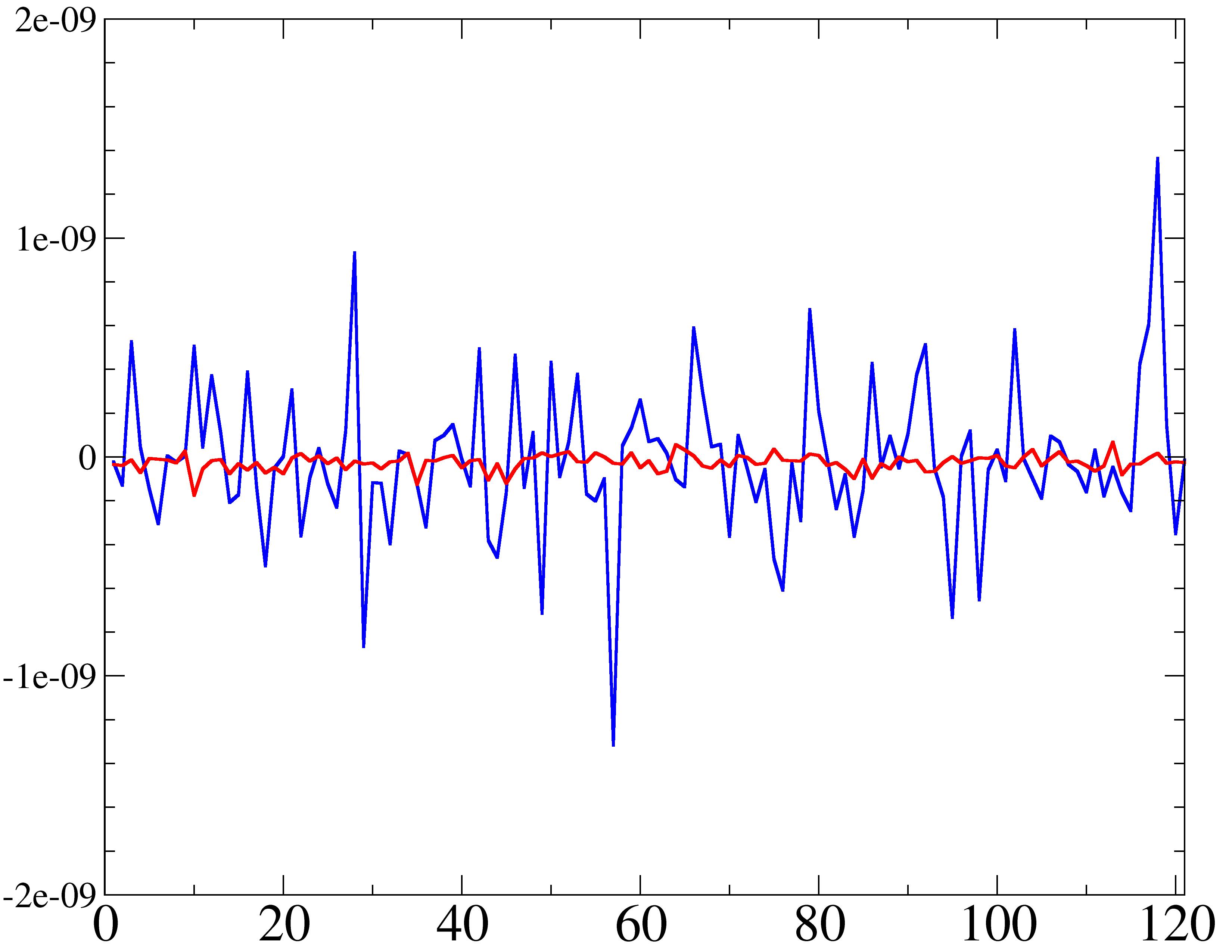}
      \end{center}
    \end{minipage}

\caption{Monte Carlo history for 120 configurations of the color-connected part of the contribution to the ratios $E_{}^{\pm}$ coming from the diagram in Figure~\ref{fig:mix3_b} for $am_c=0.04$ (left)
and $am_c=0.2$ (right).}\label{fig:mainz_MC_plot}
\end{figure}

Next we will focus on studying the performance of stochastic volume sources combined with low-mode averaging. To that purpose we will discuss results obtained on a $32\times 16^3$ lattice, again at $\beta=5.8485$. This lattice is already large enough to allow for realistic physics studies. We simulate a light quark mass of $am_{u}^{}=0.02$ together with charm masses of $am_{c}^{}=0.04,0.2$ which, using the value $a/r_{0}^{}\simeq 0.237$ from~\cite{Necco:2001xg} and $r_0=0.5$~fm, correspond to $m_{\pi}^{}\approx 318$~MeV and $m_{c}^{}\approx 64$ and $318$~MeV, respectively. Twenty low-modes are computed for each of the 120 quenched configurations. Stochastic volume sources are diluted in time, spin and color.
Out of the sixteen contributions within the LMA setup (cf. eq.~(\ref{eqn:lma_3p})), we will study in more detail a restricted subset, which either contains the main sources of statistical noise in the computation or are suitable benchmark cases. They will be labeled mix1, mix2, hlhl and lllh, and are illustrated in Figures~\ref{fig:mix1}-\ref{fig:mix3_b}. Some of them are actually the sum of various contributions; a detailed description is provided below.
For each of the contributions we will construct the ratios with products of two-point functions as per eqs.~(\ref{eqn:ratio_res_I},\ref{eqn:ratio_res_II}), and fit them to plateaux in the region in Euclidean time where the asymptotic behavior is expected to be dominated by the relevant $K\to\pi$ matrix element.\footnote{Note that, while a plateau is expected for the ratio involving the full three-point function, it is not guaranteed that the same happens for each of the LMA contributions. On the other hand, in this case the various contributions do exhibit plateaux (within the relatively large errors still present in the computation), which allows to average over an interval of time slices in order to reduce the uncertainty.} The results of these fits are shown in the lower panel of Figures~\ref{fig:mix1}-\ref{fig:mix3_b}.
The contribution mix1 contains the three diagrams with three low-mode propagators and one high-mode propagator, placed in one of the three legs; Figure~\ref{fig:mix1} shows one of them.
The data points labeled LMA make use of an extended propagator containing the leg propagator, which implies that it is possible to average over the position of the four-fermion operator. Therefore, mix1 serves as a benchmark for the efficiency of LMA+SVS for the leg propagator, since no improvement is expected when using stochastic estimates instead of the exact extended propagator. The computations reveal that stochastic all-to-all estimates computed within our setup lead to a precision comparable to the one obtained with exact extended propagators.
The contribution mix2, illustrated by the diagram of Figure~\ref{fig:mix2}, also contains the diagram obtained by switching the low-mode propagator to the right-hand side. Again, this contribution can be used as a benchmark, because the conventional LMA allows to average over space at the four-fermion operator insertion. Also in this case the combination of LMA with SVS yields comparable errors.
The diagram in Figure~\ref{fig:mix3_a}, instead, is expected to display the advantages of the hybrid approach. In this case, using LMA in conjunction with point sources does not allow to average over the four-fermion operator insertion, since two high-mode lines meet at the origin of the loop such that at least one fixed-position propagator has to attach to it.\footnote{In general, extended propagators are not applicable to contractions in which high modes run in the loop, since
in that case there is no way to construct a suitable Dirac equation that provides a product of propagators integrated
over the insertion.} 
This does not allow to use the information contained in the all-to-all low-mode propagator in the loop. Indeed we observe that the variance is reduced significantly when SVS is applied to the high-mode propagators in the legs.
Finally, the diagram in Figure~\ref{fig:mix3_b} is again expected to exhibit a significant decrease in variance when SVS is applied to the estimation of the high-mode propagator in the loop. Indeed, we find a remarkable improvement of the noise-to-signal ratio. In particular, for the heavy charm quark mass the variance is reduced by a factor of 7-8, which turns out to be decisive to obtain a signal for the total three-point functions.
Another interesting piece of information comes from the Monte Carlo history of the correlation functions. By looking at the fluctuations in the value of a correlator at some fixed value of Euclidean times a direct impression can be drawn about variance reduction, and about the presence of large fluctuations that can spoil the good gauge average behavior. In Figure~\ref{fig:mainz_MC_plot} we show the color-connected part of the contribution to the ratios $E_{}^{\pm}$ coming from the diagram in Figure~\ref{fig:mix3_b}. The advantage of LMA+SVS compared to
 LMA alone is clearly visible.
%

%
%
\subsubsection{LMA + probing versus LMA + SVS}\label{sbsbsct:SVS_EYE_DIAGRAMS_INITIAL_RUNS}
%

%
Now we will consider the hybrid strategy in which the probing approach is used to approximate the diagonal of the high-mode propagator occurring in the contribution associated with Figure~\ref{fig:mix3_b}.
Results will be quoted for the $16^4$ lattice, again at $am_{u}^{}=0.02$ and $am_{c}^{}=0.04$.
Probing is implemented with $p=2,3,4,5$. In this lattice volume this implies a total of 276,432,1452, and 2100 inversions, respectively; the factor of 12 accounting for the internal spin-color structure is included. Its results are compared to those obtained via stochastic volume sources. Dilution is applied in time, spin and spacetime (even-odd dilution) which implies a total of $\{64,128,192,256,320,640,960\}$ inversions for $\{1,2,3,4,5,10,15\}$ hits, respectively.
\begin{figure}[t!]
\centering
\subfigure[]{
\includegraphics[width=7.5cm]{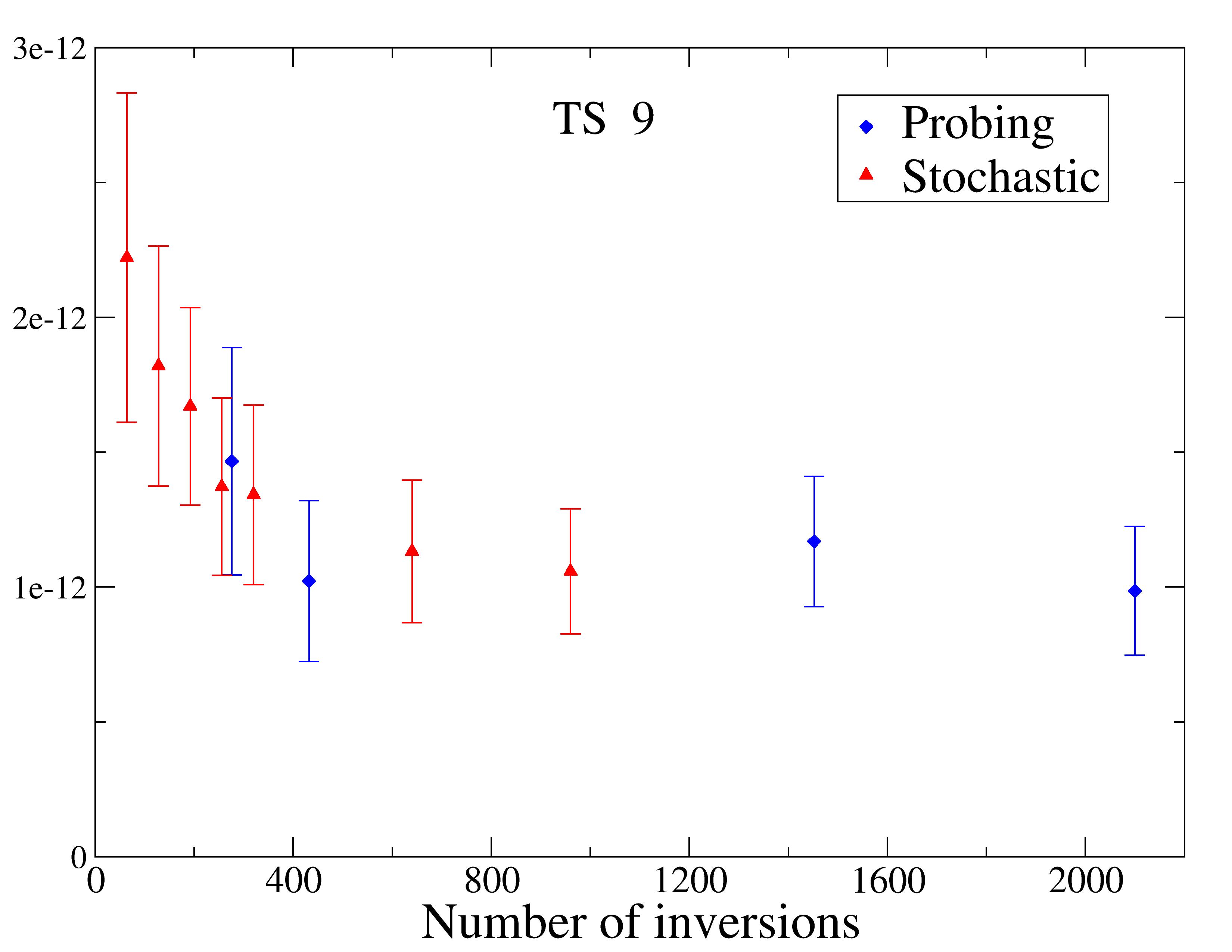}\label{fig:ERROR_SVS_VS_PROBING_CON}
}
-\hspace{-5mm}
\subfigure[]{
\includegraphics[width=7.5cm]{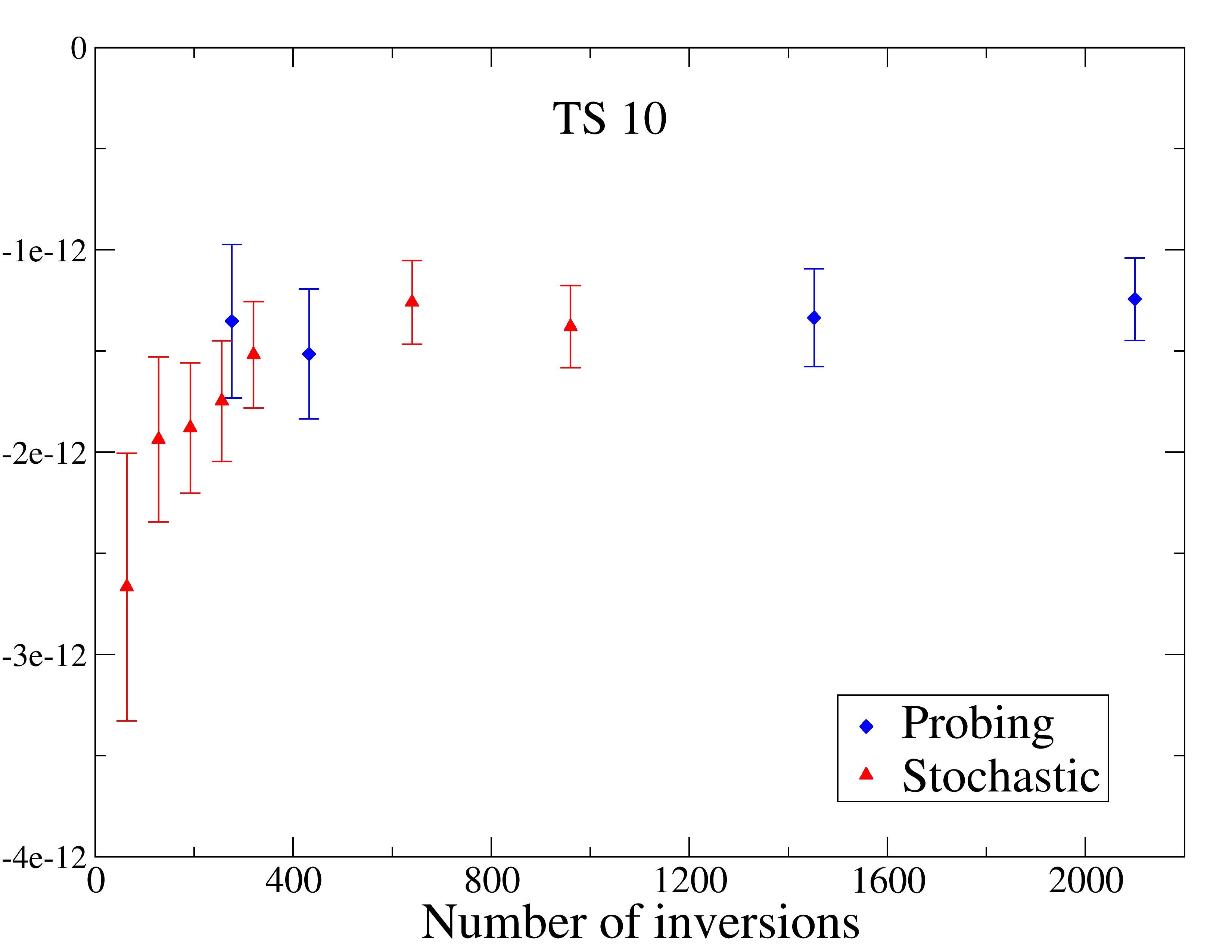}\label{fig:ERROR_SVS_VS_PROBING_DIS}
}
\caption{Probing (blue diamonds for $p=2,3,4,5$) versus SVS (red triangles for $N_{r}^{}=1,2,3,4,5,10,15$ hits) as a function of computational cost. Shown are results for the color-connected~(a) and disconnected~(b) contribution associated with the diagram in Figure~\ref{fig:mix3_b} at fixed Euclidean times $y_{0}^{}=4a,x_{0}^{}=9a$ (left) and $y_{0}^{}=4a,x_{0}^{}=10a$ (right).}
\end{figure}
In order to compare the efficiency of both methods, the results at some fixed Euclidean times are monitored as a function of the number of inversions. 
The evolution of the computed values is shown for the color-connected and color-disconnected contributions in Figure~\ref{fig:ERROR_SVS_VS_PROBING_CON} and Figure~\ref{fig:ERROR_SVS_VS_PROBING_DIS}, respectively.
Roughly $N_{r}^{}=5$ hits are required for the signal to converge in the stochastic approach. Once saturated there is an excellent agreement between the two techniques, while the errors are comparable at similar computational cost. In the color-disconnected channel, the choice $p=2$ already yields a reliable result. Increasing the number of inversions is neither justified nor necessary, since the variance only decreases moderately while the value is stable; gauge noise dominates. In the color-connected sector $p=2$ is not enough to saturate the variance, and $p \geq 3$ is required. An important observation is that in both cases the signal is saturated for $p=3$; this is of particular importance since moving from $p=3$ to $p=4$ involves a substantial increase in the number of inversions which might turn out to be prohibitively expensive for many practical applications.
While being comparable at the level of computational cost measured in the number of inversions, probing has two important advantages from a practical point of view regarding large-scale simulations. First, it results in a considerable speed-up in the contraction time, i.e. the time necessary to compute physical observables out of the propagators and appropriately chosen $\gamma$-matrices.
Note that contractions make up a non-negligible share of the total computational time in our framework,
due to the large number of vector scalar products involved: if stochastic sources are used,
the loop propagator has to be computed by performing scalar products of source and solutions vectors,
the number of which scales combinatorially with the number of hits and the number of degrees of
freedom involved in the dilution scheme; while in the case of probing the result is directly contained in just one vector.
Furthermore, and sometimes more relevant, the large memory requirements associated to stochastic
estimation (which requires availability of several vectors at any given time) for moderately large lattices rapidly give rise
to a complex storage problem, which is greatly alleviated in the case of probing.
Second, probing implies a substantial reduction in the required disk and memory space. Both issues are a
consequence of the fact that (independently on the value of the probing parameter $p$) only a single diagonal is approximated, which can be stored in just one vector. Therefore, the required memory space and contraction time are reduced to the level of standard point-to-all propagators such that for the application at hand, for instance, the overall computing time decreases up to twenty percent.
%

\section{Summary and conclusions}\label{sct:conclusion}

In this work we have explored a number of all-to-all propagator techniques in Lattice QCD:
low-mode averaging, stochastic volume sources, and probing.
Our main aim is to optimize the signal-to-noise ratio for correlation functions appearing
in the computation of non-leptonic kaon decay amplitudes; this in turn requires a detailed
study of the computation of closed quark propagators.
Our study of the probing algorithm for the computation of $S(x,x)$ is the first application
of the latter to Lattice QCD, alongside with Ref.~\cite{new_probing}, where a more sophisticated
version of the algorithm, better adapted to the structure of lattice Dirac operators, is proposed.
We have shown that the use of probing has good potential in physics applications that require
the computation of closed quark propagators; in particular, it can improve the variance reduction
attained by the use of stochastic volume sources.
We have developed a sophisticated hybrid variance reduction strategy combining low-mode averaging,
stochastic volume sources and/or probing to tackle the noise-to-signal problems of the eye diagrams
involved in the computation of $K\to\pi$ transition amplitudes.
Estimating the loop propagator in terms of all-to-all techniques results in a substantial
decrease of the variance compared to the combination of low-mode averaging with point-to-all propagators,
which is instrumental to obtain a well-behaved signal in physics applications.
In this specific case, probing does not result in a larger variance reduction than the one
attained by stochastic methods, but it does have significant practical advantages in the
computation of some contributions, resulting in faster contractions of propagators into observables and
a significant decrease in the required storage space.
Furthermore, experience with stochastic sources shows that their efficiency decreases significantly when multiple propagators have to be replaced by stochastic estimates. Using the probing method to compute the loop propagator reduces the cases where multiple stochastic propagators are necessary.
This can be used to further improve the signal.
Physics results obtained with the tools developed here are discussed in the companion paper~\cite{physpaper}.
\section*{Acknowledgments}\label{sct:acknowledgments}

E.E. and K.S. thank the kind hospitality of the Trinity College Dublin and IFT-UAM/CSIC groups,
respectively. Useful discussions with Mike Peardon are gratefully acknowledged.
Our numerical computations have been carried out at the
Altamira and MareNostrum installations of the Spanish Supercomputation Network,
the Hydra cluster at IFT, and the DiRAC facility at Edinburgh;
the authors thankfully acknowledge the expertise and technical support offered
by the staff of these centers.
This work has
been supported by the Spanish MICINN under grants FPA2009-08785 and ACI2009-1050, the Spanish MINECO under grant
FPA2012-31686 and the ÓCentro de excelencia Severo Ochoa ProgramÓ SEV-2012-0249, the
Community of Madrid under grant HEPHACOS S2009/ESP-1473, and especially by the European Union
under the Marie Curie-ITN Program STRONGnet, grant PITN-GA-2009-238353.
%

\newline


\vspace{10mm}

\begin{thebibliography}{99}

\bibitem{Bernardson}
S.~Bernardson, P.~McCarty and C.~Thron,
Comput.\ Phys.\ Commun.\  {\bf 78} (1993) 256.

\bibitem{Dong_Liu}
S.J.~Dong and K.F.~Liu,
Phys.\ Lett.\  B {\bf 328} (1994) 130, hep-lat/9308015.

\bibitem{RomeII}
G.M.~de Divitiis, R.~Frezzotti, M.~Masetti and R.~Petronzio,
Phys.\ Lett.\  B {\bf 382} (1996) 393, hep-lat/9603020.

\bibitem{Peisa}
C.~Michael and J.~Peisa  [UKQCD Collaboration],
Phys.\ Rev.\  D {\bf 58} (1998) 034506, hep-lat/9802015.

\bibitem{Foster}
M.~Foster and C.~Michael  [UKQCD Collaboration],
Phys.\ Rev.\  D {\bf 59} (1999) 074503, hep-lat/9810021.

\bibitem{SESAM}
T.~Struckmann {\it et al.}  [SESAM-T$\chi$L Collaboration],
Phys.\ Rev.\  D {\bf 63} (2001) 074503, hep-lat/0010005.

\bibitem{Cais}
A.~O'Cais, K.J.~Juge, M.J.~Peardon, S.M.~Ryan and J.I.~Skullerud
[TrinLat Collaboration],
Nucl. Phys. B (Proc. Suppl.) {\bf 140} (2005) 844, hep-lat/0409069.

\bibitem{Foley}
J.~Foley, K.J.~Juge, A.~O'Cais, M.~Peardon, S.M.~Ryan and J.I.~Skullerud,
Comput.\ Phys.\ Commun.\  {\bf 172} (2005) 145, hep-lat/0505023.

\bibitem{McNeile}
C.~McNeile and C.~Michael  [UKQCD Collaboration],
Phys.\ Rev.\  D {\bf 73} (2006) 074506, hep-lat/0603007.

\bibitem{Simula}
R.~Frezzotti, V.~Lubicz and S.~Simula,
Phys.\ Rev.\  D {\bf 79} (2009) 074506, hep-lat/0812.4042.

\bibitem{andreas_article}
P.A.~Boyle, A.~J\"uttner, C.~Kelly and R.D.~Kenway,
JHEP {\bf 08} (2008) 086, hep-lat/0804.1501.

\bibitem{ETM}
P.~Boucaud {\it et al.}  [ETM collaboration],
Comput.\ Phys.\ Commun.\  {\bf 179} (2008) 695, hep-lat/0803.0224.

\bibitem{ETMeta}
K.~Jansen, C.~Michael and C.~Urbach  [ETM Collaboration],
Eur.\ Phys.\ J.\  C {\bf 58} (2008) 261, hep-lat/0804.3871.

\bibitem{Saad}
J.M.~Tang and Y.~Saad,
Numer. Linear Algebra Appl.  {\bf 19} (2012) 485.

\bibitem{DDSaad}
J.M.~Tang and Y.~Saad,
SIAM J. Sci. Comput., {\bf 33(5)} (2011) 2823

\bibitem{LMA_DeGrand}
T.~DeGrand and S.~Schaefer, Comput. Phys. Commun. {\bf 159} (2004) 185, hep-lat/0401011.

\bibitem{LMA_hartmut}
L.~Giusti, P.~Hern\'{a}ndez, M.~Laine, P.~Weisz and H.~Wittig, JHEP {\bf 04} (2004) 013, hep-lat/0402002.

\bibitem{Hartmut} 
L.~Giusti, P.~Hern\'{a}ndez, M.~Laine, P.~Weisz and H.~Wittig, 
JHEP {\bf 11} (2004) 016, hep-lat/0407007.

\bibitem{prl}
L.~Giusti, P.~Hern\'andez, M.~Laine, C.~Pena, J.~Wennekers and H.~Wittig,
  Phys.\ Rev.\ Lett.\  {\bf 98} (2007) 082003.
  
\bibitem{zm}  
P.~Hern\'andez, M.~Laine, C.~Pena, E.~Torr\'o, J.~Wennekers and H.~Wittig,
  JHEP {\bf 0805} (2008) 043.
  
\bibitem{physpaper}
E.~Endress and C.~Pena,
  Phys.\ Rev.\ D {\bf 90} (2014) 9,  094504,
  arXiv:1402.0827 [hep-lat].

\bibitem{proceedings}
E.~Endress and C.~Pena,
PoS ConfinementX {\bf } (2012) 110;
PoS LATTICE {\bf 131} (2012).

\bibitem{Blum:2012uh}
  T.~Blum, T.~Izubuchi and E.~Shintani,
  Phys.\ Rev.\ D {\bf 88} (2013) 9,  094503
  [arXiv:1208.4349 [hep-lat]].

\bibitem{new_probing}
A.~Stathopoulos, J.~Laeuchli and K.~Orginos,
SIAM J. Sci. Comput. {\bf 35} (2013) S299,
hep-lat/1302.4018.

\bibitem{Bali:2009hu}
  G.~S.~Bali, S.~Collins and A.~Schafer,
  Comput.\ Phys.\ Commun.\  {\bf 181} (2010) 1570
  [arXiv:0910.3970 [hep-lat]].

\bibitem{Morningstar:2011ka}
  C.~Morningstar, J.~Bulava, J.~Foley, K.~J.~Juge, D.~Lenkner, M.~Peardon and C.~H.~Wong,
  Phys.\ Rev.\ D {\bf 83} (2011) 114505
  [arXiv:1104.3870 [hep-lat]].

\bibitem{Alexandrou:2013wca}
  C.~Alexandrou, M.~Constantinou, V.~Drach, K.~Hadjiyiannakou, K.~Jansen, G.~Koutsou, A.~Strelchenko and A.~Vaquero,
  Comput.\ Phys.\ Commun.\  {\bf 185} (2014) 1370
  [arXiv:1309.2256 [hep-lat]].

\bibitem{bump}
L.~Giusti, M.~L\"uscher, P.~Weisz and H.~Wittig, JHEP {\bf 11} (2003) 023, hep-lat/0309189.

\bibitem{Bernardoni:2010nf}
  F.~Bernardoni, P.~Hern\'andez, N.~Garron, S.~Necco and C.~Pena,
  Phys.\ Rev.\ D {\bf 83} (2011) 054503
  [arXiv:1008.1870 [hep-lat]].
    
\bibitem{SVS_Wilcox}
W.~Wilcox, (1999), hep-lat/9911013.

\bibitem{SVS_BULAVA}
J.~Bulava, R.~Edwards and C.~Morningstar,
PoS LATTICE {\bf 125} (2008), hep-lat/0810.1469.

\bibitem{GRAPH_COLORING_3}
A.R.~Curtis, M.J.D.~Powell and J.K.~Reid,
J. Inst. Math. Appl. {\bf 13} (1974) 117.

\bibitem{GRAPH_COLORING_2}
T.F.~Coleman and J.J.~More,
SIAM J. Numer. Anal. {\bf 20} (1983) 187.

\bibitem{GRAPH_COLORING_1}
T.F.~Coleman,  B.S.~Garbow and J.J.~More,
ACM Trans. Math. Software {\bf 10} (1984) 346.
	
\bibitem{GRAPH_COLORING_4}
A.H.~Gebremedhin, F.~Manne and A.~Pothen,
SIAM Rev. {\bf 47} (2005) 62.
	
\bibitem{Bekas}
C.~Bekas, E.~Kokiopoulou and Y.~Saad,
Applied Numerical Mathematics {\bf 57} (2007) 1214.

\bibitem{SAAD_BIBLE}
Y.~Saad,
Iterative Methods for Sparse Linear Systems, 2nd edition, (2003), 
Society for Industrial and Applied Mathematics SIAM, Philadelphia.

\bibitem{DDHMC}
M.~L\"{u}scher, 
Comput. Phys. Commun. {\bf 165} (2005) 199, hep-lat/0409106;
JHEP {\bf 12} (2007) 011, hep-lat/0710.5417.

\bibitem{DD}
Y.~Saad, Iterative methods for sparse linear systems, 2nd
edition (SIAM, 2003);\\ 
A.~Quarteroni and A.~Valli, Domain
decomposition methods for partial differential equations
(Oxford University Press, 1999).

\bibitem{HMC}
S.~Duane, A.D.~Kennedy, B.J.~Pendleton and D.~Roweth,
Phys. Lett. B {\bf 195} (1987) 216.

\bibitem{Sexton}
J.C.~Sexton and D.H.~Weingarten,
Nucl. Phys. B {\bf 380} (1992) 665.

\bibitem{SAP_GCR}
M.~L\"{u}scher,
Comput. Phys. Commun. {\bf 156} (2004) 209, hep-lat/0310048.

\bibitem{procsdublin}
  L.~Giusti, C.~Pena, P.~Hern\'andez, M.~Laine, J.~Wennekers and H.~Wittig,
  PoS LAT {\bf 2005} (2006) 344.
  
\bibitem{unpublished}
  L.~Giusti, C.~Pena, P.~Hern\'andez, J.~Wennekers and H.~Wittig, unpublished.
  
\bibitem{noise_rbc}
D.~Zhang, {\em Using all-to-all propagators for ${K}\to\pi\pi$ decays},
Talk presented at the 31th International Symposium on Lattice Field Theory, Mainz, Germany (2013).
  
\bibitem{OVERLAPP_1}
H.~Neuberger,
Phys. Lett. B {\bf 417} (1998) 141, hep-lat/9707022.

\bibitem{OVERLAPP_2}
H.~Neuberger,
Phys. Lett. B {\bf 427} (1998) 353, hep-lat/980103.
  
\bibitem{Necco:2001xg}
S.~Necco and R.~Sommer, 
Nucl. Phys. B {\bf 622} (2002) 328, hep-lat/0108008.	

\bibitem{oveps}
  L.~Giusti, C.~Hoelbling, M.~L\"uscher and H.~Wittig,
  Comput.\ Phys.\ Commun.\  {\bf 153} (2003) 31.
  
\end{thebibliography}
\end{document}